\documentclass[journal]{vgtc}                % final (journal style)
\ifpdf%                                % if we use pdflatex
  \pdfoutput=1\relax                   % create PDFs from pdfLaTeX
  \usepackage{graphicx}                % allow us to embed graphics files
  \DeclareGraphicsExtensions{.pdf,.png,.jpg,.jpeg} % for pdflatex we expect .pdf, .png, or .jpg files
\else%                                 % else we use pure latex
  \usepackage{graphicx}                % allow us to embed graphics files
  \DeclareGraphicsExtensions{.eps}     % for pure latex we expect eps files
\fi%

\graphicspath{{figures/}{pictures/}{images/}{./}} % where to search for the images

\usepackage{microtype}                 % use micro-typography (slightly more compact, better to read)
\PassOptionsToPackage{warn}{textcomp}  % to address font issues with \textrightarrow
\usepackage{textcomp}                  % use better special symbols
\usepackage{mathptmx}                  % use matching math font
\usepackage{times}                     % we use Times as the main font
\usepackage{cite}                      % needed to automatically sort the references
\usepackage{tabu}                      % only used for the table example
\usepackage{tabularx}
\usepackage{changepage}
\usepackage{booktabs}                  % only used for the table example
\usepackage{enumitem}
\setlist[itemize]{noitemsep, topsep=0pt}
\usepackage{wrapfig}
\usepackage{caption}
\usepackage{titlecaps}
\usepackage{dblfloatfix}
\usepackage{amsmath}
\usepackage{amssymb}
\usepackage{newtxmath}
\usepackage{fontawesome}
\usepackage{placeins}
\usepackage{color}
\usepackage{physunits}

\usepackage{etoolbox}
\AtBeginEnvironment{quote}{\par\singlespacing}

\definecolor{cb_orange}{rgb}{1.0,0.51,0.0}
\definecolor{cb_blue}{rgb}{0.22,0.49,0.72}
\definecolor{cb_green}{rgb}{0.3,0.67,0.29}
\definecolor{cb_red}{rgb}{0.89,0.1,0.11}
\definecolor{cb_purple}{rgb}{0.6, 0.31, 0.64}
\definecolor{cb_brown}{rgb}{0.6, 0.4, 0.2}
\definecolor{cb_crimson}{rgb}{0.86, 0.08, 0.24}

\newcommand{\re}[1]{{\textcolor{black}{#1}}}
\newcommand{\para}[1]{\vspace{1mm}\noindent\textbf{#1}}

\def\summary{\texttt{Summary\:Mode}}
\def\game{\texttt{Game\:Mode}}

\onlineid{1103}

\vgtccategory{Research}
\vgtcpapertype{Application/Design Study}

\title{VIRD: Immersive Match Video Analysis for \\High-Performance Badminton Coaching}
\author{Tica Lin$^{1,2}$, Alexandre Aouididi$^{1,3}$, Zhutian Chen$^1$, Johanna Beyer$^1$, Hanspeter Pfister$^1$, Jui-Hsien Wang$^2$}

\authorfooter{
\item $^1$ Harvard John A. Paulson School of Engineering and Applied Sciences

\item $^2$ Adobe Research
\item $^3$ The École polytechnique fédérale de Lausanne (EPFL)
}

\abstract{
Badminton is a fast-paced sport that requires a strategic combination of spatial, temporal, and technical tactics. 
To gain a competitive edge at high-level competitions, badminton professionals frequently analyze match videos to gain insights and develop game strategies. 
However, the current process for analyzing matches is time-consuming and relies heavily on manual note-taking, due to the lack of automatic data collection and appropriate visualization tools. 
 As a result, there is a gap in effectively analyzing matches and communicating insights among badminton coaches and players. 
This work proposes an end-to-end immersive match analysis pipeline designed in close collaboration with badminton professionals, including Olympic and national coaches and players. 
We present \emph{VIRD}, a VR Bird (i.e., shuttle) immersive analysis tool, that supports interactive badminton game analysis in an immersive environment based on 3D reconstructed game views of the match video.
We propose a top-down analytic workflow that allows users to seamlessly move from a high-level match overview to a detailed game view of individual rallies and shots, using situated 3D visualizations and video. 
We collect 3D spatial and dynamic shot data and player poses with computer vision models and visualize them in VR. Through immersive visualizations, coaches can interactively analyze situated spatial data (player positions, poses, and shot trajectories) with flexible viewpoints while navigating between shots and rallies effectively with embodied interaction. 
We evaluated the usefulness of VIRD with Olympic and national-level coaches and players in real matches. Results show that immersive analytics supports effective badminton match analysis with reduced context-switching costs and enhances spatial understanding with a high sense of presence.

} % end of abstract

\CCScatlist{ % not used in journal version
 \CCScat{K.6.1}{Management of Computing and Information Systems}%
{Project and People Management}{Life Cycle};
 \CCScat{K.7.m}{The Computing Profession}{Miscellaneous}{Ethics}
}

\teaser{
 \hspace{-1cm}
 \includegraphics[width=1.1\linewidth]{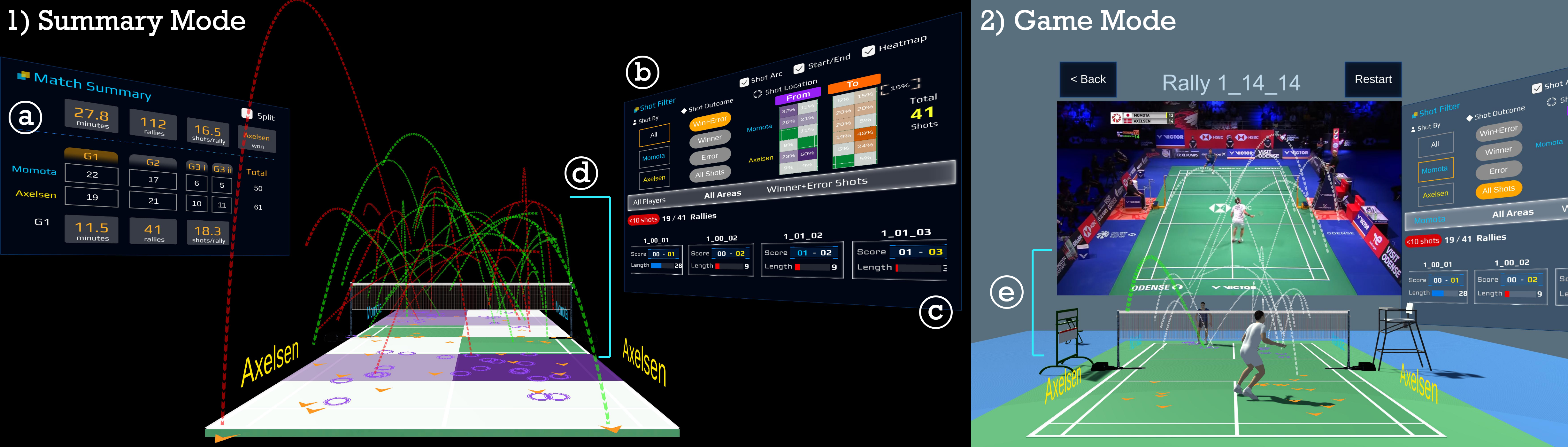}
 \caption{
 VIRD is an immersive VR platform for top-down badminton match analysis. 
 Left: Users start with a high-level Match Summary (a), then refine their analysis using the Shot Filter (b). Detailed rally and shot information is available through the Rally Menu (c) and Situated Visualizations (d) on a virtual court. 
 Right: Users can link to a Game View (e) of a selected shot (\summary{}) or an entire rally (\game{}), featuring synchronized video and 3D dynamic player and shot representations.
 }
    \label{fig:teaser}
}

\vgtcinsertpkg
\begin{document}


% !TEX root = ../main.tex

\firstsection{Introduction}
\maketitle

% Professional badminton players compete at the highest level, 
% and the outcomes of their matches are often decided by the narrowest of margins. 
% Players and coaches have come to rely heavily on match analysis to identify key patterns and strategies that can be used to gain a competitive advantage against their opponents. 
In the highly competitive world of professional badminton, 
coaches and players constantly seek ways to gain an edge over their opponents. 
Detailed match analysis has become indispensable for identifying key patterns, 
developing winning strategies, 
and creating tailored training plans. 
Traditional methods of match analysis require coaches 
to review video footage of matches,
take notes, and identify trends and potential areas for improvement.

%%%% talk about why the traditional methods are bad
%%% "the coach needs to context switch a lot between video and spatial data" + "ineffective communication using the current medium".
However, these traditional methods have significant drawbacks.
% First, traditional methods require coaches to manually annotate videos to collect in-game data about the players,
% such as their statistical summaries, 
% distributions of the shuttle placements,
% and trajectory types.
% Coaches often have to watch countless rallies multiple times, limiting the time they can spend with their players. 
% 
First, traditional methods require coaches to manually annotate videos while analyzing the match. In order to gather insights from match footage, coaches have to watch videos several times to collect data, such as players' statistical summaries, shuttle locations and types, and playing styles. After that, coaches have to compare the summary data with their observations to iterate on their hypotheses, verify insights, and draw conclusions.
This data collection and analysis process can be highly time-consuming and mentally demanding, as coaches have to watch countless rallies multiple times and maintain a high level of attention to detail. While the analysis efforts vary widely, a badminton match can contain 50 to over 100 rallies and the coaches typically spend 3 to 5 hours analyzing a full match video according to our formative study.
% \jui{might be good to cite some real numbers here. How many rallies, how much time they have to spend wrt to video length} 
As a result, this can increase their cognitive load and limit the time coaches spend working with their players.
% 
% Coaches may spend three to five times the duration of a match collecting statistics from crucial game moments, 
% cross-referencing data and video to verify insights, 
% and communicating their advice to the players.
Furthermore, traditional methods struggle to capture and communicate badminton's inherently spatial nature effectively. 
Badminton players use height- and speed-varying tactics to control the pace and aggression of their games, which is harder to perceive in projected 2D videos.
In turn, coaches rely on physical demonstrations to help players conceptualize the insights. The limited coaching time available for high-performance athletes can restrict the amount of information conveyed during coaching sessions.
% 
% this can lead to missing insights or communicating difficulty \jui{can we support this claim? it looks weak and hand wavy}.
% crucial to effective match analysis and coaching.
% Thus, players often require seeing actual movement to understand the concepts. 
% Traditional medium also limits the coach's ability to fully convey these spatial aspects to players.
Finally, traditional methods separate the in-game data from its physical context.
Consequently, coaches often need to switch back and forth between the videos and the collected data,
inevitably increasing the mental effort of the analysis process.
As shown in our formative studies (Sec.~\ref{sec:formative_study}), there is a demand for more efficient match analysis and coaching tools.

In this work, 
we closely collaborated with Olympic badminton professionals
to design VR Bird (VIRD)\footnote{A badminton shuttlecock is also informally called a \emph{bird}.}, a novel interactive coaching tool for badminton match analysis in VR.
To address the limitations of traditional methods, VIRD adopts a top-down analysis approach to support an efficient iterative analysis process. It leverages computer vision (CV) techniques to extract game statistics and 3D shot and player data, and enables situated analysis of spatial and dynamic game data in immersive 3D space with an interactive approach. 
\re{Immersive analytics were found beneficial in supporting 3D data analysis with spatial understanding and immersion~\cite{t_chandler_immersive_2015}.
With its high portability and affordability, 
we chose VR HMD as our targeted platform to design immersive analysis solutions for badminton video match analysis.}

To design VIRD, we performed a design study to answer the following three research questions.
First, we conduct a formative study with badminton professionals to understand 
\textit{``What data are required for analyzing matches and developing coaching insights?''} 
Second, we identified gaps and iterated solutions with coaches to answer
\textit{``What is the ideal coaching workflow for badminton video analysis and communication?''}. 
Finally, we conducted a multi-staged user-centered design to address \textit{``How to design an integrated video analysis tool to support badminton coaching for professional coaches and players?''}
We conducted case studies with high-performance badminton experts, including Olympic and national team coaches and players, on match analysis using VIRD. Both coaches and the player were able to perform effective match analyses, and verify and present insights using VIRD with high satisfaction. They leveraged immersive 3D visualizations to generate new insights with concrete evidence, such as observing shot distributions or pinpointing specific game moments, and used VR interaction to accelerate their iteration from hypotheses to insights.  
% leveraging CV-based data collection and immersive analysis approach to accelerate their analysis process. 
% \begin{itemize}
% \item What data are required for analyzing matches and developing coaching insights? 
% \item What is the ideal coaching workflow for badminton video analysis and communication?
% \item How to design an integrated video analysis tool to support badminton coaching for professional coaches and players?
% \end{itemize}

% \jui{Another question I have is whether we can elevate the framing of the paper to include other competitive individual sports like Tennis, and then specify why and how we focus on badminton.}

Our research has four main contributions:
1) a formative study with Olympic badminton professionals to identify gaps in current match analysis workflow for coaching, 
2) a characterization of goals and tasks for badminton match analysis in coaching, 
3) an end-to-end immersive video analysis tool, VIRD, with state-of-the-art CV-based data collection and a VR analytic interface for badminton coaching,
and 4) case studies with high-performance badminton experts to evaluate the usefulness of VIRD.
Our results suggest that applying immersive analytics to sports videos, with CV-based data collection and human-in-the-loop analysis, can be highly effective for sports professionals.

\section{Related Work}

%\subsection{Visual Analytics of Sports Games}
\noindent
\textbf{Visual Analytics of Sports Games.}
\re{
Sports visualization research often aims to visualize spatial and dynamic data in context, e.g., on a court diagram (basketball\cite{pingali2001visualization}, tennis\cite{wu2018forvizor}, baseball\cite{dietrich2014baseball4d}), embedded in the game videos (soccer~\cite{stein2017bring}, basketball~\cite{lin2022quest, chen2023iball}) or projected on a virtual court in immersive environments (baseball~\cite{zou_evaluation_2019}, badminton\cite{chu2021tivee}).}
This is largely because contextual understanding is crucial for deriving meaningful and actionable insights from sports games~\cite{tuyls2021game,patton2021predicting,heaton2023perform}.
%There has been a long interest in sports visualization research to visualize spatial and dynamic data in context, e.g., on a court diagram (basketball\cite{pingali2001visualization}, tennis\cite{wu2018forvizor}, baseball\cite{dietrich2014baseball4d}), embedded in the game videos (soccer~\cite{stein2017bring}, basketball~\cite{lin2022quest, chen2023iball}) or projected on a virtual court in immersive environments (baseball~\cite{zou_evaluation_2019}, badminton\cite{chu2021tivee}).
%This is largely because contextual understanding is crucial for understanding sports games and for deriving meaningful and actionable insights~\cite{shea_2014,tuyls2021game,patton2021predicting,heaton2023perform}.

For racket sports in particular, spatial and temporal data, such as shot locations and trajectories, are crucial for game analysis and regulations.
LucentVision~\cite{pingali2000lucentvision} is a commercial tennis visualization system based on real-time ball/player tracking. It offers virtual replays of ball trajectories and presents a color heatmap to show the coverage of player movements.
%They designed several novel visualizations such as a color heatmap on a court diagram showing the coverage of player movements, and a virtual replay of a detected ball trajectory from flexible angles.
Similarly, the Hawk-eye system~\cite{owens2003hawk} 
provides 3D views of tracked tennis balls using multiple cameras, and extends to other sports like baseball and soccer for enhanced game viewing and officiating.
% provides 3D views of the tracked tennis ball based on multiple tracking cameras and has been expanded to other sports, such as baseball pitches and soccer goals, where spatial visualizations are shown in virtual game views to enhance game viewing and officiating.

Two prior studies used immersive analytics to  analyze spatial badminton stroke data. 
ShuttleSpace~\cite{ye2020shuttlespace} visualizes badminton shot trajectories in a VR environment to support coaches in analyzing shot data from a first-person perspective.
 They provide an integrated visual design that augments 3D trajectory data with 2D statistical information using a first-person perspective visualization and peripheral vision. ShuttleSpace also enables natural and efficient trajectory selection in VR with a stroke metaphor, allowing analysts to select trajectories by imitating badminton strokes. The system has been evaluated through case studies conducted by domain experts, demonstrating its potential for facilitating badminton data analysis.
\re{TIVEE~\cite{chu2021tivee} designed an immersive VR system for experts to analyze sequential stroke trajectories in badminton.  It allows experts to explore different tactics from a third-person perspective, and provides a detailed court view for inspecting and explaining tactics that lead to wins and losses. Case studies with professional badminton experts demonstrate the system's effectiveness in identifying patterns of commonly used tactics.}

\re{
In contrast to tennis, official badminton games do not use tracking systems. Thus, 
most badminton professionals lack access to manually collected shot datasets~\cite{ye2020shuttlespace, chu2021tivee} and rely on
video-based match analysis (Sec.~\ref{sec:formativestudy}).
Our study fills this gap by providing an end-to-end immersive video analytic tool, VIRD, that integrates computer vision-based data collection directly from match videos. Compared to prior work~\cite{chu2021tivee,ye2020shuttlespace}, our study provides comprehensive match analysis for coaching, emphasizing both analysis and communication of insights.
}
% While prior work~\cite{chu2021tivee,ye2020shuttlespace} concentrate on immersive visualization of static stroke trajectories, VIRD provides a dynamic and immersive 3D representation of complete badminton match data, including situated visualizations and 3D shot and player models. 
% Our study emphasizes  a comprehensive approach to support match analysis for coaching, both in analyzing and communicating insights.

% Our study focuses on match video analysis for badminton coaches and visualizes dynamic shot and player data based on automatic data collection generalizable to all match videos (Sec.~\ref{sec:data_preprocessing}). 

%\subsection{3D Game Reconstruction of Sports Videos}
\para{Game Reconstruction of Sports Videos.}
Reconstructing 3D sports games from videos offers opportunities to improve game understanding and analytics. Computer vision research has 
focused on reconstructing game scenes and player or ball movement for various sports, such as basketball~\cite{chen2009physics, zhu2020reconstructing}, tennis~\cite{pingali2000lucentvision,owens2003hawk}, soccer~\cite{rematas2018soccer}, volley ball~\cite{chen20113d}, or human motion in sports~\cite{rematas2018soccer}.
Badminton games present unique challenges,
such as single-camera recordings, fast shuttle speed (the fastest shuttle speed can be over 250 mph), and complex player movement.
% Among all sports, badminton games present several unique challenges. First, badminton
% games are captured by single-camera videos, as usually no multi-angled or tracking cameras are available in tournaments. Second, tracking the shuttlecock and reconstructing its trajectory can be difficult due to its fast speed and unique physics. The fastest shuttle speed can be over 250 mph. Third, reconstructing 3D player movement on the court is not trivial due to the required accuracy on player shapes and poses.
% Third, reconstructing 3D player movement on the court is not trivial due to the required accuracy on player shapes and poses.
% Two main challenges exist. First, tracking the small shuttlecock and reconstructing its trajectory can be difficult due to its fast speed and unique physics. The second challenge is to accurately detect and reconstruct the player's poses and shape on the court from the single-angled video footage.

\re{Prior studies have addressed shuttlecock detection and tracking, including an instant review system to determine whether a shot was in or out~\cite{kopania-2022}.
%Kopania et al.\cite{kopania-2022} developed an instant review system that determines whether a shot was in or out.
} 
Other studies focused on tracking the speed, rotation angle, and athlete's body transformation using sensors and path-tracking algorithms\cite{lyu-2021}. Recently, MonoTrack~\cite{liu-2022} improved state-of-the-art models~\cite{huang2019tracknet, farin2003robust,wang2020deep} on court recognition and 2D trajectory estimation based on badminton domain knowledge. \re{
We use MonoTrack~\cite{liu-2022} to accurately extract and segment 3D shuttle trajectories in match videos.
%This method accurately extracts and segments 3D shuttle trajectories, providing a more precise analysis of badminton games. We use MonoTrack~\cite{liu-2022} to reconstruct dynamic shots in the match videos.
}

While there has been some research on the detection and tracking of badminton players \cite{haq-2022,rahmad-2019}, there has been relatively little focus on estimating their poses and shapes. Fortunately, SMPL~\cite{SMPL:2015},
a learned model of human body shape, allows accurate human shape representation from basic human 3D models. 
In addition, recent state-of-the-art algorithms are capable of performing 3D human pose and shape estimation \cite{zou-2021, monet-2022}. Among them, CLIFF~\cite{li-2022} can estimate SMPL parameters from a 2D video, which we use to reconstruct player poses.

% Our study used MonoTrack~\cite{liu-2022} to reconstruct dynamic shots and track player positions, which were then combined with pose estimation by CLIFF~\cite{li-2022} to reconstruct player poses in the badminton game.

%\subsection{Immersive Analytics for Spatial and Dynamic Data}
\para{Immersive Analytics for Spatial and Dynamic Data.}
Immersive analytics (IA) has gained significant attention among visualization researchers due to its ability to facilitate analytical reasoning and collaboration for analyzing high-dimensional and multivariate data~\cite{t_chandler_immersive_2015, 
% cordeil_imaxes:_2017,
% marriott_immersive_2018, 
ens2021grand}. 
IA offers several advantages over traditional desktop visualizations~\cite{kraus2022immersive} due to its large screen spaces and embodied interaction offered by AR/VR technologies: immersion~\cite{millais2018exploring, helbig2014concept}, multi-modal interaction~\cite{butscher2018clusters, lopez2015towards, ready2018immersive}, and presenting data in-situ~\cite{benko2004collaborative, hachet2011toucheo, moran2015improving}, which improves spatial understanding and user experience throughout the visual analysis process.
%
%With its large screen spaces and embodied interaction offered by immersive technologies, such as AR and VR displays, IA offers several advantages over traditional desktop visualizations~\cite{kraus2022immersive}, such as immersion~\cite{millais2018exploring, helbig2014concept}, multi-modal interaction~\cite{butscher2018clusters, lopez2015towards, ready2018immersive}, and presenting data in-situ~\cite{benko2004collaborative, hachet2011toucheo, moran2015improving}, which improves spatial understanding and user experience throughout the visual analysis process.
% 
% However, shortcomings of IA exist in certain steps in the analytic workflow, such as annotation tasks that require precise manipulation~\cite{dube2019text}, or displaying statistical and abstract information~\cite{munzner2014visualization}.
% 
Specifically, prior studies found several benefits of analyzing spatial and dynamic data in immersive environments, i.e., identifying spatial attributes such as distance~\cite{yang2018maps, kraus2019impact}, transitioning between 2D and 3D views~\cite{yang2020tilt}, and the visceral experience of viewing dynamic, physical data in VR~\cite{lee2020data}.

With these identified benefits of IA, recent works investigate applying immersive analytics to sports~\cite{lin2020sportsxr}. Besides ShuttleSpace~\cite{ye2020shuttlespace} and TIVEE~\cite{chu2021tivee} for badminton shuttle trajectory analysis in VR, 
Lin et al.~\cite{lin2021} present real-time basketball shot arcs for situated analysis in AR during free-throw training. Sumiya et al.~\cite{sumiya2022anywhere} applied a similar approach for basketball shooting training in VR. 
Rezzil~\cite{rezzil} provides a commercial soccer training system in VR that allows players to evaluate their performance with instant visual feedback. Zou et al.~\cite{zou_evaluation_2019} presents real-time bat swing spatial data for baseball batting training in VR.

% With more spatial and dynamic data involved in data analysis, 
% recent research proposed immersive analytics tool kits to allow authoring and analyzing spatial data
% \cite{buschel2021miria, sicat2018dxr, lee2023deimos, cordeil2019iatk}, or integrated analysis methods of 3D and 2D data with mixed reality ~\cite{hubenschmid2022relive, cavallo2019immersive}.
% These works establish the groundwork for domain-agnostic data exploration and analysis, such as observing user interaction in a lab study.
%While many previous work applied 
% IA has been applied to scientific visualization fields, such as archaeology~\cite{benko2004collaborative, smith2013artifactvis2} and biomedicine~\cite{nowke2013visnest, maes2018minomics}. 
% Few IA works address domain-specific analysis tasks for non-scientific users, such as sports~\cite{lin2020sportsxr, chu2021tivee}, IoT~\cite{ens2017ivy}, and facility management~\cite{prouzeau2020corsican, coupry2021bim}. 

While most IA work in sports focuses on providing real-time feedback for training, our study focuses on immersive video analytics for sports coaching with spatial and dynamic data extracted from sports videos. To the best of our knowledge, we are the first study to propose an immersive analysis system for sports video analysis.

% We aim to empirically evaluate the benefits of applying IA in sports and tackle some of the challenges outlined by Ens et al.~\cite{ens2021grand}, such as how situated contexts benefit users' cognitive load.

% !TEX root = ../main.tex

\section{Formative Study with Olympian Coaches \& Players}
\label{sec:formative_study}
We applied a user-centered design process to develop VIRD and involved target users at every design stage. All experts involved in our study are Olympic or national team coaches and players.
Section \ref{sec:formative_study} presents the gaps in match analysis based on expert interviews with coaches and players.
Section \ref{sec:goal_task_analysis} presents our goal and task analysis, which informed the design of VIRD.
Section \ref{sec:vird} presents VIRD's design based on three rounds of user testing with coaches.
Section \ref{sec:user-study} presents the evaluation of VIRD with both coaches and players on match analysis for developing game strategy and communicating insights. 
% Note that due to the specific domain we target, i.e., high-performance badminton coaching, our design is guided by a small number of domain experts involved in the study. While this is the nature of professional sports, we discuss the implications of conducting research with professional athletes in Sec.~\ref{sec:proathletes}.

% \jui{unprofessional question here: should this section really be called "Design Requirement"? Doesn't seem to match what comes next... But maybe it is a field thing..}

% We interviewed 5 badminton professionals on their coaching experiences to obtain insights into gaps in their current video analysis. 
% We iterated with 3 coaches to propose an ideal match analysis workflow and conducted goal and task analysis to form design requirements. 

% \jui{One thing I think will be good to clarify here or in intro is why we have so few interviewees -- due to our focus on high level players.}

\subsection{Procedures}
\label{sec:formativestudy}
To understand current practice and identify gaps in badminton coaching, we interviewed five professional badminton players and coaches (I1-I5; M = 2, F = 3; Age:
30-45). All of them are former Olympic players representing Canada, Taiwan, and the US. All of them have at least 10 years of player experience and four became professional coaches after their playing careers with 2 to 15 years of coaching experience.

We conducted 1-hour semi-structured interviews online to elicit the interviewee's background, overall coaching workflow, video and data usage in their coaching, and how they evaluate the player's performance in the video analysis. Finally, we asked interviewees to analyze a short match video to demonstrate their typical analysis workflow. 

All interviews were transcribed and analyzed using affinity mapping. Our analysis focused on understanding the current badminton coaching practice and identifying gaps in their match analysis workflow.

%%%
\subsection{Findings and Gaps}
\label{sec:findings}
Overall, we observed that coaching practices varied widely among the interviewees due to varying resource levels, player skill levels, and coaching styles. 
%Despite differences in coaching style, 
The coaching process typically involves a significant amount of video analysis for both coaches and players. 
% why videos
For players, 
videos are crucial as they help players become aware of their playing technique and style and allow them to create a mental model of other players.
Players are often told to record their own match and watch the videos multiple times, 
e.g., \textit{``Most coaches recommend we watch it several times and break it down to focus on one thing at a time''} (I3). 
Coaches also rely on videos to direct coaching by analyzing the root cause of player performance and
communicating insights to players,
e.g., \textit{``They won’t understand what I am saying unless they see it physically''} (I2).

% This practice is useful for developing strategic insights into the game. 

% Therefore, analyzing matches for coaching is crucial for both coaches and players.
All interviewees agreed that video analysis is time-consuming. 
I4 noted that
\textit{``if the match is 30-40 minutes, it took 3 to 5 hours to rewatch and discuss with your coach''}. 
When analyzing the videos, 
coaches and players watch matches, take notes, analyze them for insights, and discuss their findings. 
With more resources, coaches can proactively share their insights with players, but this requires a significant time investment (I2). 
Alternatively, in cases where access to a coach is limited, 
players may seek coaching by requesting feedback on areas to improve (I3, I5)
or by coaching each other (I1, I3, I4). 
Therefore, analyzing matches for coaching is crucial for both coaches and players.

For clarity on badminton terminology, note that a match comprises the best-of-three games. Each game has multiple rallies, with each rally awarding a point. Within every rally, players execute a series of shots.
Below, we summarize four gaps in the current match video analysis.

% Coaches found it beneficial to analyze multiple matches of their players over a period of time (I1, I2, I4).
% \textit{``If you had like 20 matches that you played for an entire year, then you have a better idea where you can dictate training''} (I1). 

% This variation can be explained by a few variables. First, the level of resources available to a player can vary significantly depending on their level of competition, with high performers or older players typically receiving more attention. 
% Second, the coach-player relationship 
% can impact the dynamics and ultimately the outcome of the coaching, as emphasized by a coach that \textit{``you really have to understand the individual and know how they respond to certain things''} (I2). 
% Third, coaches' own preference and familiarity with the available technology stack shape their coaching style. 
% Some coaches use data analysis (e.g., notational analysis) to reveal errors and shot patterns (I1, I2), while others gain insights on playing styles and shot quality from match videos (I3, I4).
% Finally, the player's skill level impacts the instruction types. 
% Novices receive guidance on basic tactics and pose correctness, while elite athletes focus on tactical-level instruction.
% In this work, we summarized the current practice and pain points in video analysis.

\subsubsection{The Current Bottom-Up Workflow is Inefficient}
\label{sec:gap1}

We observed that the experts we interviewed followed a \emph{bottom-up} approach to generate their insights. 
They began by scanning videos to detect insights into a player's playing style and weaknesses.
For example, during video scanning, I3 promptly observed, \emph{``She's a lefty ... [so] she tends to lean more to her left side for a big forehand.''} 
Upon forming an insight hypothesis, they scrutinized additional videos to identify similar patterns and validate their observations, e.g., \textit{``Is this just an outlier or some of these random matches where we didn’t do well, or concrete things that we need to work on?''} (I2). 
Coaches repeatedly watched games until insights emerged and revisited the games to gather further evidence.
Because of the limited time and resources to review footage, coaches and players might choose to only look at the most important parts, which leads to incomplete analysis
and potentially less effective communication. 
In summary, the current bottom-up analysis workflow lacks support for efficient iteration and analytic reasoning.
% \zt{I feel we need a sentence to transit to the top-down workflow. maybe quote ``Overview first, details on demand''}
% leading us to the first design goal:

\subsubsection{Manual Data Collection from Videos is Time-Consuming}
\label{sec:gap2}

During the current workflow,
experts manually collect summary statistics from watching the videos to reveal patterns quantitatively. This allows them to compare the player performance (I1, I2, I4) and
communicate better with the players (I1, I2).
For example, 
having these data benefit their coaching greatly, e.g., \emph{``If you had like 20 matches that you played for an entire year,
then you have a better idea where you can dictate training''} (I1).
Further, concrete evidence like \textit{``70\% of time when you do this, you win the point''} (I2) allows players to immediately grasp the concept.
% 
% Players also compare their performance in tournaments and training by counting patterns from selected rallies, such as shot types, locations, sequence, and body movement.
However, 
when watching a video,
paying attention to multiple metrics and patterns simultaneously is difficult. 
As a result, coaches and players have to watch the videos multiple times and
focus on different aspects one at a time, 
such as opponent versus their player,
and winner shots versus unforced errors. 
Such a manual process \emph{``takes up a lot of time''} (I2). In summary, manually collecting data from videos hinders experts to perform match analysis efficiently.
% This manual process is time-consuming and might result in incomplete analysis and less effective communication. Therefore, it is crucial to incorporate a design goal that addresses this issue.

\subsubsection{Data Insights and Contexts are Presented Separately}
% \para{Presenting Data Insights with Videos is Necessary}
\label{sec:gap3}
% \zt{need presenting insights with videos}

Currently, data insights and videos are often presented separately. Coaches typically provide players with a summary of key insights, such as 
\textit{``you are pushing the tempo and make too many mistakes (6 errors in 11 points you lost)''},
without directly connecting to specific moments in the video. 
This disjointed presentation hinders a comprehensive understanding of the game, as players may struggle to visualize the context behind the numbers. 
To bridge this gap, coaches might manually note timestamps of critical game moments with the help of some tools (e.g., YouTube, Hudl~\cite{hudl}, Clutch~\cite{clutch}). However, collecting the video moments is still very tedious, as noted by I2 that \textit{``I have to spend 30 minutes per player writing things down, and another 1 hour to review notes with them to show them here’s what happened''}.
Moreover, without ample video evidence, sometimes it can be hard to convince the players of a particular finding. For example, I1 mentioned that 
\textit{``The kids don't realize that they make a lot of unforced errors. If they watched the video clips, they'd understand it better''}.
Therefore, the current way of presenting data and video separately may impede a holistic understanding of the game due to an inefficient workflow.

\subsubsection{2D Game Representation is Insufficient}
\label{sec:gap4}

When asked about the limitation of analyzing matches with videos, 
multiple coaches expressed that single-camera recordings might fail to capture essential game aspects, 
such as environmental conditions (e.g., wind), 
shot timing (\textit{``Speed seems slower in the video''}), 
and player reactions.
Although official badminton match are limited by monocular videos,
some coaches use 360-degree videos in training for comparing players and their opposition, as 
\textit{``you can see both sides and like 
how the player reacts to the opponent in real-time''} (I1). 
Slow-motion or zoomed views also assist in breaking down techniques and offering objective perspectives (I1, I3).
I3 noted \textit{``when you're hitting the shot, you only know how it feels, but you can't see how it looks''}. 
These remarks indicate that traditional 2D videos fall short in providing spatial comprehension and flexible viewing angles necessary for analyzing specific shot attributes. %monocular
This finding aligns with previous work~\cite{ye2020shuttlespace, chu2021tivee}.

\subsection{Summary}

Based on the formative study, we identified that high-performance badminton
coaches and players perform match analysis to reveal the
strengths and weaknesses of players and to develop playing
or training strategies. However, data collection (i.e., bottom-up workflow and manual note-taking) and presentation (i.e., separation of data and videos and 2D game representations) gaps exist in their current analysis workflow, leading to inefficient match analysis and communication for coaching.

% !TEX root = ../main.tex
\section{Goal \& Task Analysis} 
\label{sec:goal_task_analysis}

\subsection{Design Goals}
To support coaches and players in analyzing matches and communicating
insights effectively, we characterized four design goals with respect to the four identified gaps in  Sec.~\ref{sec:findings}.

\para{G1. Providing a top-down analysis workflow.} 
Our tool needs to present summary data to enable an immediate overview of the match, and support effective data exploration to discover regions of interest for detailed analysis. 
Unlike the traditional bottom-up approach that requires users to watch the videos sequentially to observe insights, the top-down approach supports users to analyze game details on demand, driven by observed patterns from summary data,
\re{as captured in ``Overview first, zoom and filter,
details on demand''~\cite{shneiderman2003eyes}}.
% Per coaches' feedback, essential metadata include scores, rally count, match duration, errors and winners count, and length of each rally. 

\para{G2. Collecting data from videos automatically.}
To avoid tedious manual data collection from users, our tool must provide the necessary data, including summary statistics and annotation of critical shots (i.e., winners and errors). This data should be automatically collected without user input during the match analysis.
% \jui{The goal says it needs to be "automatic", but our preprocessing is not automatic.}
   
\para{G3. Integrating abstract data with game contexts.} 
% \zt{merge abstract data with the physical context} 
% to support comprehensive analysis and communication.} 
Even though summary data can help coaches identify patterns in the game, 
it is important to investigate the actual game moments in the video to verify observations and analyze root causes, as well as to present the insights to players.
\re{Similar to the concept of "Search, show context, expand on demand” on large graphs~\cite{van2009search}, the user goal is to search for a meaningful context.}
Our tool should provide an easy transition between statistics and videos to support iterative analytical reasoning and communication.

\para{G4. Visualizing spatial data in 3D space.} 
3D data should be 
visualized within a 3D space to support an accurate interpretation of 
their spatial attributes, e.g., shot speed and trajectory. 
In addition, our visualizations should support an analysis from flexible viewpoints to enhance users' spatial perception and support objective perspectives.

\subsection{Task Abstraction} 
\label{sec:task_abstraction}
We abstract six analytic tasks users perform when analyzing matches. Currently, users have to manually gather summary data to identify rallies of interest and manually navigate to each rally to extract insights.

\para{T1. Identify the rallies of interest based on game summary data.}
 % \zt{a}
 % A user first needs a game breakdown to obtain an overview of the match (e.g., game length, winning rallies). Then, they often want to filter a set of interesting rallies based on the metadata of the games.
Users first obtain an overview of the match performance at the game level (e.g., game length, scores).
Based on the game metadata, they can focus on a subset of interesting rallies, such as the winning rallies by their player in the first game.
% \textit{Gap: There is currently no direct way to filter and navigate to rallies based on game information.}

\para{T2. Identify the rallies of interest based on rally summary data.}
 % \zt{b}
%  In addition to the metadata of the game,
% the user often wants to obtain an overall impression of the rallies (e.g., playing style, rhythm) from a shot level,
% such as the ratio of winner shots and distribution of the shot locations.
% Such statistics allow them to further filter rallies with specific patterns for the next step of analysis.
After filtering rallies based on the game summary, users observe patterns of the game (e.g., playing style, competition) at the rally level,
such as the ratio of winner versus error rallies.
Such statistics allow them to identify specific patterns and further filter rallies for deeper analysis.
% \textit{Gap: Currently, users have to watch the entire video and manually collect this data.}
% - Shots and rallies are not categorized and have to be found manually.  

% Identify patterns of winner and error shots.
\para{T3. Gain a statistical overview of rallies of interest.}
 % \zt{c}
Focusing on a set of filtered rallies, users compare the statistics of individual rallies to obtain high-level insights, such as assessing the pace of a rally from shot counts and stress levels from score differences. 
% the user then focuses on analyzing the winner and error shots of each rally to derive insights. 
% This requires a dashboard to show the summary statistics for each rally.
% \textit{Gap: Users have to manually count data for each rally with the current tools.}

\para{T4. Gain a spatial overview of rallies of interest.}
 % \zt{d}
 Additionally, users examine spatial and temporal aspects of individual shots among these rallies (e.g., shot trajectory and speed) and compare them against rally summary data (e.g., shot location distributions).
The spatial information can help users gain an in-depth understanding of the shot-level data and form insights on specific rallies.
% Additionally, 
% showing the spatial and temporal aspects of the shots (e.g., trajectories and shuttle speed) and their comparison with the summarized attributes (e.g., shot types and placement distributions)
% in space is essential
% for badminton analysis.
% The spatial information can help the users can gain an in-depth understanding of the data and derive more insights.
% \textit{Gap: Current methods do not present spatial and dynamic data in 3D space. }

\para{T5.  Investigate game details of specific shots.}
 % \zt{e}
 \re{
 After gaining an overview, users can dive into rallies to examine game details, such as player movement and shot sequence, to get deeper insights into a player's performance.
 This often requires users to watch the game moment multiple times from different angles (e.g., player vs. opponent).}
% Once some insights are observed, 
%the user dives into the respective game moments of the rally to examine game details missing from the summary statistics, such as player movement and shot sequence, to verify the observations and dive deeper into the cause of the performance. 
%This often requires the user to watch the game moment multiple times from different angles (e.g., player vs. opponent).
% Once some patterns are observed, 
% the user dives into the respective game moments of the rally to examine game details missing from the summary statistics, such as player movement and strategy, to contemplate deeper into the cause of the performance. 
% This often requires the users to watch the game moment from different angles.
% \textit{Gap: When watching a video, the users cannot interpret dynamic shots and player movement from different angles.}

%%%% shot level
\para{T6.  Verify insights across rallies.} 
 % \zt{c}
Before concluding their analysis, users need to cross-validate other rallies with similar or contrasting patterns to update and verify their insights.
Thus, users need to efficiently navigate to other game moments based on observed patterns.
% \textit{Gap: With a mainstream video player, it is tedious to navigate and replay specific game moments based on their data patterns.}
% Navigating and replaying specific game moments in the video is tedious, which causes high mental load to compare summary statistics with game details. }
% - It is time-consuming to compare and organize video clips to support insights.

% !TEX root = ../main.tex

\section{VIRD - \underline{V}R B\underline{ird} Video Analysis Tool}
\label{sec:vird}
\re{We designed VIRD, our immersive video analysis platform for high-performance badminton coaching, targeting professional badminton coaches and players. 
VIRD features a top-down analysis approach and supports an integrated data and video analysis workflow based on CV-based data collection and a 3D interactive environment in VR. We iterated the designs based on expert feedback from three coaches.}
%Based on the identified goals and tasks, we designed VIRD, an immersive video analysis platform for high-performance badminton coaching. Targeting professional badminton coaches and players, VIRD features a top-down analysis approach and supports an integrated data and video analysis workflow based on CV-based data collection and a 3D interactive environment in VR. We iterated the designs based on expert feedback from three coaches.
% The code for VIRD is open-source\footnote{The code for VIRD will be made available at https://to-be-open.github.io}.

\subsection{Top-Down Analysis Workflow} 
 To address G1, we designed a \emph{top-down analysis approach} and verified it with two coaches (I1 and I2) in a follow-up interview.
 The top-down user flow contains four steps: First, users find rallies of interest based on summary statistics (\textbf{T1, T2}). Second, they compare and analyze the filtered rallies to gain initial insights (\textbf{T3, T4}). Third, they investigate game details to extract more specific insights (\textbf{T5}). Finally, they examine similar patterns across rallies to verify their insights (\textbf{T6}).
 
We tested the proposed workflow
by gathering coaches' feedback on a hypothetical top-down analysis workflow for examining a lost game:

\begin{adjustwidth}{0.5cm}{0.5cm}
Initially, the user reviews the game summary to form an impression of the match. 
 Observing a tight score of 17 (\textit{Player A}) to 21 (\textit{Player B}), the user chooses to analyze the 21 rallies where \textit{Player A} lost points (\textbf{T1}). 
They find 10 errors among the 21 lost points and focus their analysis on those 10 rallies ending in error shots (\textbf{T2}). 
On the rally level, the user finds that 6 out of 10 error rallies are short and towards the end of the game (\textbf{T3}).
Examining the heat map and shot trajectories, the user notes that 70\% are from the middle, 
and identifies them as defensive shots (\textbf{T4}). 
The user selects a short rally (\textbf{T5}) to study player movement and the sequence leading to the error shot. 
After watching multiple error rallies  (\textbf{T6}), the user concludes that \textit{Player A} needs to improve defensive shots on the backhand side and work on physical fitness.
\end{adjustwidth}

\normalsize
Both coaches agreed that this workflow is precisely what they need. \textit{``We're manually doing that because currently I won't know where exactly to go back in the video to see all the unforced errors''} (I2).

\subsection{Computer Vision Data Preprocessing}
\label{sec:data_preprocessing}
To address G2, we applied state-of-the-art computer vision (CV) models to automate the data collection from videos.
We developed a semi-automatic data processing pipeline for monocular match videos, which includes a manual game breakdown, shot classification algorithms, and automatic 3D shot and player reconstructions.
Based on our task abstraction, three types of data must be extracted from a match video:

\para{1) Game and Rally Summary} are automatically computed based on manual annotation and output from CV models:

\begin{itemize}[leftmargin=*]
  \item \emph{Rally Breakdown}: 
    To obtain a game summary, player scores and aggregated rally statistics are required. Therefore, each game needs to be split into rallies. 
    We manually annotated the time ranges (start and end), player who serves, and the winning side of all rallies from the match video. Our algorithm then derives score and game information based on the rally breakdown.

    \item \emph{Shot Breakdown}: 
    To obtain a rally summary, the duration of each rally and shot count are required. 
    We obtain the timestamps and each shot's hitter running MonoTrack~\cite{liu-2022} with minor manual clean-up.
\end{itemize}

\para{2) 3D Spatial Data} are automatically reconstructed from CV models.
\begin{itemize}[leftmargin=*]
    \item \emph{3D Shot Trajectory}: 
    We automatically reconstruct 3D trajectories and velocities for all shots with MonoTrack~\cite{liu-2022}.

    \item \emph{3D Player Model}: 
    To reconstruct 3D player models, we use MonoTrack~\cite{liu-2022} to estimate court and player positions. We use CLIFF~\cite{li-2022} to predict smooth 3D player poses from videos. %This combination allows for the reconstruction of 3D player models.
\end{itemize}

\para{3) Shot Statistics} are automatically derived from the 3D spatial data based on experts' analysis requirements gathered in the formative study. 
% \jui{I think if we can show some helpful illustrations for these derived statistics, it can be useful} 
\begin{itemize}[leftmargin=*]
    \item \emph{Shot Tendency}: 
    To detect whether a shot leading to a point is a winner or an unforced error, we first classify the shot tendency by approximating the shuttle’s velocity vector when it passes the net: the tendency is defensive when the vector is going upward (away from the ground), and offensive if opposite.  %(as described in \emph{Shot Outcome})
    % This classification is derived from expert input in the formative study.
% \jui{this is motivated by Toby's need right? Should we say this (if not here, then somewhere)?}
% The velocity information is obtained from running MonoTrack~\cite{liu-2022}.

    \item \emph{Shot Outcome}:
% \jui{There are two problems here: 1) the writing makes it confusing what is shot tendency and what is classification. One way to separate them is: tendency is the attempt of the shot, and the "classification" is the outcome based on this attempt. I will clarify these two paragraphs more. 2) "classification" might not be the right choice. When I read "shot classification", I immediately think of the types of the shot, like "clears", "drives", "net drops", "smashes" etc.}
    Our algorithm categorizes shots as winners, errors, or normal shots to calculate the winner and error shot counts.
    A rally ends with a winner by the scorer or an error by the point loser.
    If the last shot is offensive by the scorer, it's a winner; if defensive by the point loser, the penultimate shot is the winner. Conversely, if the last shot is offensive by the point loser, it's an error; if defensive by the scorer, the penultimate shot is an error. All others are normal shots.
% % winner
% If the final shot is an offensive shot hit by the scorer, then it is labeled as a winner for the scorer.
% If the final shot is a defensive shot hit by the point loser, then the second last shot is a winner for the scorer.
% % error
% Conversely, if the final shot is an offensive shot hit by the point loser, then it is labeled as an error; if it is a defensive shot hit by the scorer, then the second last shot is labeled as an error by the point loser.

    \item \emph{Shot Distribution}:
    From the formative study, coaches use shot locations to classify shots into six areas on the court, including front/middle/back on the left and right sides. 
    To compute the shot distribution, 
    our algorithm projects each shot's start (from) and endpoint (to) onto the court to decide shot locations.

% Our algorithm projects each shot's start and end point onto the court\jui{it might not be clear to the reader why the projection is needed. Explain it slightly more.}, and computes the shot distribution across seven court areas on each side, including front/middle/back on the left and right sides or out-of-bound.

\end{itemize}

\vspace{1mm}
\noindent
Overall, our data preprocessing pipeline is largely automated. 
Except for rally breakdown and winner/server annotation, all other data are obtained through automatic algorithms. 
% \jui{the winners and serves are also annotated. Also, I might supplement that this limitation can be addressed in cv but is out of scope for this paper}.
The manual annotation takes up roughly half of the video duration (e.g., 30 minutes for labeling a 1-hour video).
While our method is not entirely automatic, we anticipate that CV techniques may be able to address these manual annotations in the future, though they are beyond the scope of our current study.

\subsection{Visual Designs}
\label{sec:visual_components}

\begin{figure}[t]
  \centering
  \includegraphics[width=\linewidth]{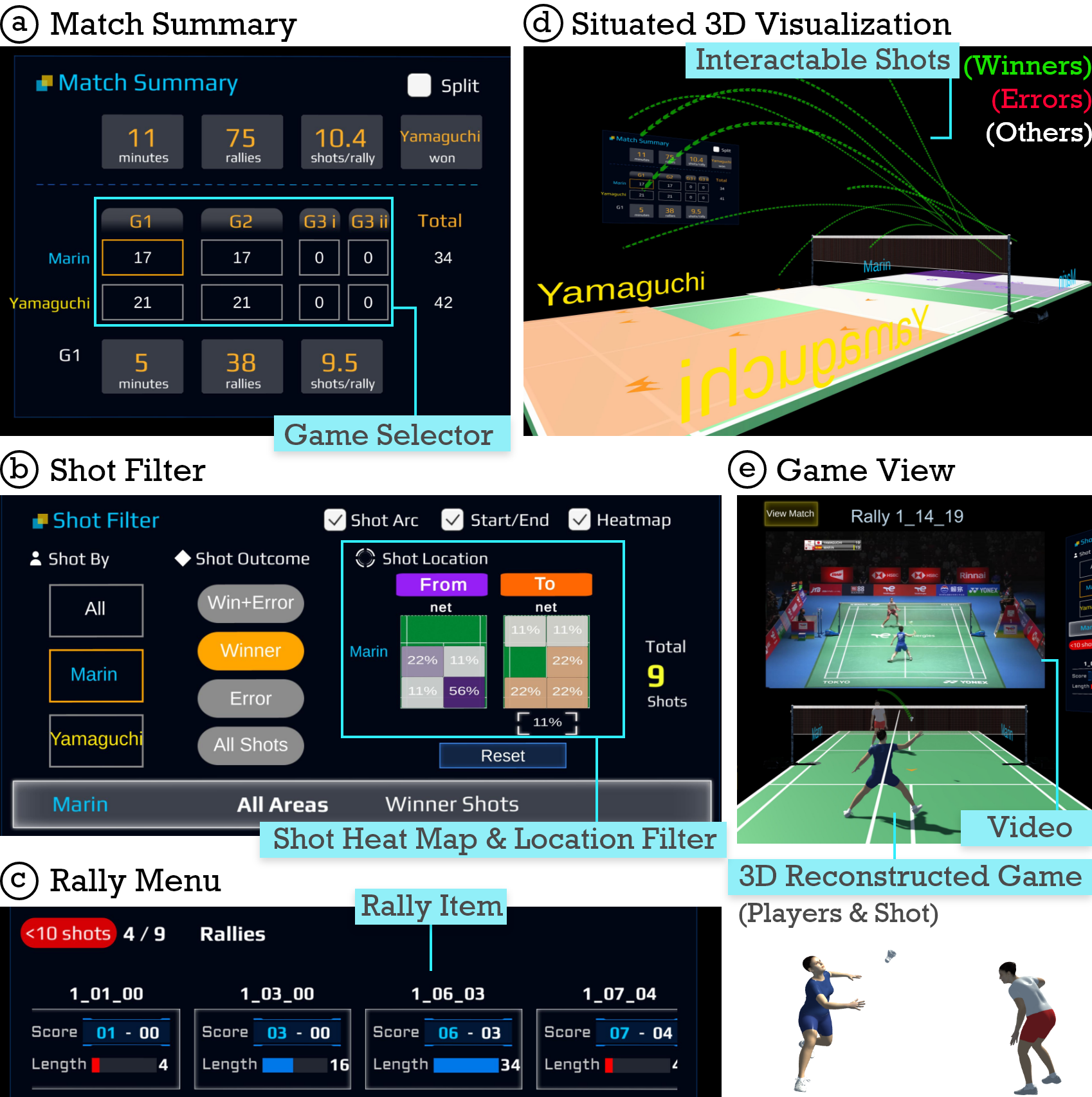}
  \vspace{-5mm}
  \caption{VIRD Visual Design.  (a) and (b) provide a match overview and allow users to examine and filter rallies listed in (c) based on game and shot details. Static immersive visualizations of 3D shot data are shown in (d), whereas dynamic game views are displayed in (e).
   }
    \vspace{-4mm}
  \label{fig:VIRD}
\end{figure}

% \subsubsection{Statistical summary and filtering}
We designed five visual components to support the users' analytic tasks and fulfill design goals (G1, G3 and G4).
On a high level, users analyze data across rallies in \summary{} (\autoref{fig:teaser}-1) and dive into a specific rally in \game{} (\autoref{fig:teaser}-2). 
We describe components with examples
% Descriptions are 
based on the 2022 BWF World Championship match between Marin and Yamaguchi~\cite{match_marin_yamaguchi} (M2) used in our case studies (\autoref{sec:user-study}). 

% In order to identify rallies of interest, experts need to break down the match and examine the statistics of games and rallies.  
% We designed Match Summary and Shot Filter to support experts getting necessary metadata and filtering data in Summary Mode.

% \setlength{\intextsep}{0pt}%
% \begin{wrapfigure}{L}{0.23\textwidth}
% 	\centering
% 	\includegraphics[width=0.23\textwidth]{images/componentA.png}
% 	   \vspace{-8mm}
% 	\caption{Match Summary}
% 	 % \vspace{-2mm}
% 	\label{fig:match_summary}
% \end{wrapfigure}

% \begin{figure}[h]
%     \centering
%     \includegraphics[width=0.5\linewidth]{images/componentA.png}
%     \vspace{-3mm}
%     \caption{Match Summary}
%     \vspace{-4mm}
%     \label{fig:match_summary}
% \end{figure}

 % Match Summary 
\para{(a) Match Summary}
(\autoref{fig:VIRD}a), supporting \textbf{T1},
enables users to identify rallies of interest
by showing essential statistics, including the match's duration, rally count, average shot count per rally, winner, and game scores. 
% These statistics are essential to overview the pace and intensity of the match. 
Users can select a game and view the rallies won by a player from the game selector (e.g., 17 rallies won by Marin in G1). 
 Similar statistics for the selected game are displayed below. 
Game 3 is split into two halves by default due to the switch of sides at the midpoint, 
which prevents spatial data from being displayed on the same side.
 These statistics help experts identify more challenging or outstanding games for deeper analysis.
\re{Furthermore, based on user feedback, we added the option to split games into first and second halves for finer granularity. This was based on the need to}
% Our formative study found that experts sometimes 
analyze each half separately due to coaching advice provided during the midpoint break.
% Game 3 is split into two halves by default, as players switch sides at midpoint (when the leading side reaches 11 points) and therefore spatial data cannot be displayed on the same side of the court.
% These metadata help experts identify games that were more challenging or outstanding for deeper analysis.
% In addition, users can `split' a game into first and second half, similar to Game 3. This design was to provide finer granularity of the game. In the formative study, we found that sometimes experts would analyze each half separately, as players can receive coaching advice in a short break at the midpoint of each game. 

% Shot Filter

% \FloatBarrier
% \begin{wrapfigure}{R}{0.3\textwidth}
% 	\centering
% 	\includegraphics[width=0.3\textwidth]{images/componentB.png}
% 	 \vspace{-8mm}
% 	\caption{Shot Filter}
% 	 % \vspace{-2mm}
% 	\label{fig:shot_filter}
% \end{wrapfigure}

% \begin{figure}[h]
%     \centering
%     \includegraphics[width=0.7\linewidth]{images/componentB.png}
%     \vspace{-3mm}
%     \caption{Shot Filter}
%     \vspace{-4mm}
%     \label{fig:shot_filter}
% \end{figure}

% Once the game is selected, 
\para{(b) Shot Filter} (\autoref{fig:VIRD}b), supporting \textbf{T2}, enables users to analyze specific shots based on player and shot attributes. 
\re{Finding specific game moments is important to analyze strengths and weaknesses. Therefore, our filter design includes key metrics to support immediate access to the necessary details, including players, shot outcomes, and locations.}
\re{Coaches in our user testing found it extremely valuable to analyze shots filtered by players.}
% Coaches in our user testing found this feature extremely valuable for comparing players. 
For instance, out of Marin's 17 points, 9 were scored through Marin's winners, while 8 were scored due to Yamaguchi's errors.
This provides a different perspective than if Marin had 17 winners. 
\re{Users can also analyze shot distribution by location filtering.}
% Users can also view shot distribution on the heatmap and filter shots by area. 
For instance, users can select the purple 56\% grid to filter Marin's winner shots from the back right. 
Darker colors on the heatmap indicate more shots are from (purple) or to (orange) the area. 
\re{We picked the color scheme to avoid visual clutter and provide an easier comparison of hot spots and shot tendencies between players and games.}

\para{(c) Rally Menu} (\autoref{fig:VIRD}c), supporting \textbf{T2} and \textbf{T6},  provides an overview of shot count and scoring cadence for the rallies of interest, with direct access to the specific rally upon selection. 
Each rally is displayed in a scrollable list with score and length information. 
Short rallies (less than 10 shots) are highlighted in red to draw special attention based on coaches' requirements. 
For instance, users can discern the game's tempo from the number of short rallies (4 out of 9) and the variation of shot counts among the rallies won by Marin. 
This overview of rally statistics helps identify patterns across rallies of interest (\textbf{T4)}.
Further, users can investigate game details in \game{} by selecting a rally (\textbf{T6)}, which allows easy transition to the game context.

% Situated Vis
% \FloatBarrier
% \begin{wrapfigure}{R}{0.25\textwidth}
% 	\centering
% 	\includegraphics[width=0.25\textwidth]{images/componentD.png}
% 	 \vspace{-6mm}
% 	\caption{Situated Visualizations}
% 	 % \vspace{-2mm}
% 	\label{fig:virtual_court}
% \end{wrapfigure}

% \begin{figure}[h]
%     \centering
%     \includegraphics[width=1\linewidth]{images/3d.pdf}
%     \vspace{-4mm}
%     \caption{Left: Situated Visualizations. Right: Game View.}
%     \vspace{-4mm}
%   \label{fig:virtual_court}
% \end{figure}

\para{(d) Situated 3D Visualizations} (\autoref{fig:VIRD}d), supporting \textbf{T4}, display the 3D shot arcs and heatmap of filtered shots (e.g., all winners by Marin in Game 1) on a 1-to-1 virtual court. 
Shots are color-coded based on their outcome, with red for errors, green for winners, and white for all other shots. 
Interacting with individual shot arcs displays the shuttle's dynamic trajectory in real-time (details in Sec.~\ref{sec:interaction}).
The situated visualizations enable users to glean insights into shots' spatial attributes, 
such as arc shapes and distributions. 
Experts use this design to quickly observe insights from shots across multiple rallies.

% \subsubsection{Investigate game details of shot and rally}
% \FloatBarrier
% \setlength{\intextsep}{0pt}%
% \begin{wrapfigure}{R}{0.25\textwidth}
% 	\centering
% 	\includegraphics[width=0.25\textwidth]{images/componentE.png}
% 	\vspace{-6mm}
% 	\caption{Game View}
% 	 % \vspace{-2mm}
% 	\label{fig:game_vieiw}
% \end{wrapfigure}
% Once some patterns are observed, it is necessary for experts to investigate actual game details to verify insights and investigate root causes. The user can preview the game moments of the chosen shot in Summary Mode, or delve into the entire rally in Game Mode.

\para{(e) Game View} (\autoref{fig:VIRD}e), supporting \textbf{T5}, shows a 3D reconstructed game view along with the video to facilitate a more comprehensive analysis of game details.
The 3D game view displays the dynamic movement of shots and players, enabling accurate spatial perception and flexible viewing angles. 
By examining the exact moment of the selected shot, users can quickly compare multiple shots in \summary{} and expand their analysis to the selected rally in \game{}.
% or to other rallies containing similar shots from the Rally Menu (\textbf{T6}).

% 
% Fig.~\ref{fig:Interaction}-2 shows the user interaction of linking from a shot in Summary Mode to the rally in Game Mode. 

% The user can select `View Match' from the hovered shot preview in Summary Mode (Fig.~\ref{fig:Interaction}-1) to watch the entire rally in Game Mode (Fig.~\ref{fig:Interaction}-2).

% 

\subsection{User Interaction}
\label{sec:interaction}

\begin{figure}[t]
  \centering
  \includegraphics[width=\linewidth]{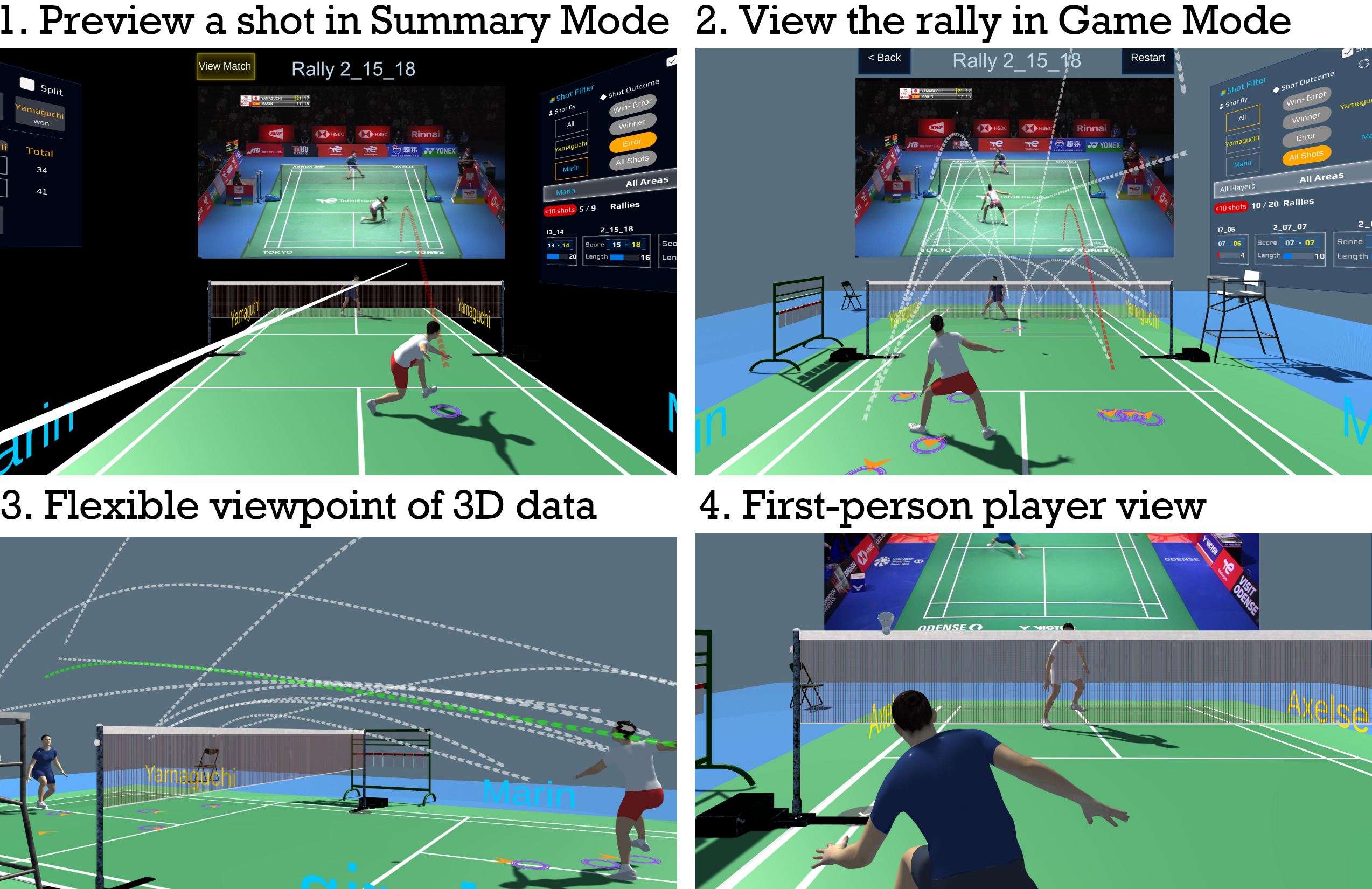}
  \vspace{-5mm}
  \caption{VIRD Interaction. 
    Users directly point at visualizations to 1) preview a hovered shot and 2) link to the entire rally. They can use a thumb stick to 3) move flexibly in VR and 4) change viewing angle.}
    \vspace{-4mm}
  \label{fig:Interaction}
\end{figure}

\noindent
Users interact with the VIRD interface and visualizations using VR controllers.
Each shot can be hovered to select, which will link to the game context of the shot (\autoref{fig:Interaction}-1) while in \summary{}, showing a Game View that contains 3D dynamic shot trajectory and player poses, and the same shot duration in the video.
This interaction allows users to instantly review the game moment of each shot in the filtered group (e.g., all winner shots) to obtain the context of 3D data. 
To navigate to the \game{} (\autoref{fig:Interaction}-2), the users can select \textit{``View Match''} from the Game View of the hovered shot, or select a rally from the Rally Menu.
Furthermore, the user can directly hover over a shot arc in the rally (\autoref{fig:Interaction}-3) to play the video from the desired game moment.
This feature allows replaying a specific shot or shot sequence efficiently.

Meanwhile, users can flexibly navigate the virtual court, by using a thumb stick or physically moving around, to obtain an accurate spatial and temporal perception of 3D data (\autoref{fig:Interaction}-3).
Our VR environment also offers flexible viewpoints to analyze the 3D game from different perspectives, such as the first-person player view (\autoref{fig:Interaction}-4).

\subsection{Design Iterations}
\label{sec:design_iteration}
We conducted three rounds of user testing throughout the design process with three active high-performance US badminton coaches (C1-C3; M=3; Age: 40-60). All were former players on the US, Malaysia, and Nepal badminton national teams, respectively. 
They have coaching experiences ranging from 15 to over 30 years. C1 had participated in our formative study while C2 and C3 were newly introduced at the design iteration stage.
Given the challenges in accessing domain experts, we adopted a progressive approach wherein each coach evaluated our prototype at various design stages, focusing on distinct aspects. 
We tested VIRD on the Women's Single match of the 2020 BWF World Tour Finals between Tai Tzu Ying and Carolina Marin~\cite{match_tai_marin}.

\para{Round 1. User flow and data analysis.} The first round of testing was conducted on the initial prototype with C1, where we elicited the coach's feedback on the overall analysis approach and the shot filtering features.
The coach appreciated the top-down approach and interactive analysis process with immediate access to all match data and videos.
On top of the existing summary data and filters, he suggested showing winner and error shots separately to support an immediate comparison of the shot patterns.
Further, \re{we designed two interaction methods to apply the shot location filter,  1) select buttons on the Shot Filter panel and 2) physically move to the desired area on the virtual court. However, the coach felt 1) is more useful as moving around the court to filter shots during the analysis would be tedious and distracting.}

\para{Round 2. Interaction with the visualization and interface.} The second round of testing was run a month later with C2, with a focus on the interaction of linking the static data to the dynamic trajectories and videos. We showed the dynamic shot trajectory of a hovered shot arc, but to view the original rally video the user had to scroll through and select from the Rally Menu.
The coach suggested augmenting the preview of the selected shot arc with the video, as \textit{``it takes time to find the video part of it right now''}.
He also emphasized the importance of pinpointing on the cause of the outcome during coaching. It is not enough to see where an error shot occurred in general, but to let the player see the shot sequence and player movement that lead to the outcome. 
Therefore, we implemented a shot-to-rally interaction (\autoref{fig:Interaction}-1 to 2), which allows coaches to look into a specific rally (and the cause for errors/winners) more efficiently. 
%where the error/winner occurs for detailed analysis into root cause. 

\para{Round 3. Immersive 3D visualization.} We conducted the third test two months later with C3 on the usefulness of immersive visualizations. The coach was able to preview each group of filtered shots (e.g., winners) efficiently and interpreted the shot types from the 3D shot arcs to answer his coaching questions, like \textit{``What are the shots Tai used to win?''- 1 cross drop [shot], 2 smash [shots], 1 block [shot]}. 
He also valued the color usage in the visualizations to tell the shot percentage on the heat map and highlight winner and error shots.
However, since 3D player poses were not implemented at the moment, we found that coaches still watched the video view to analyze the rally, as player movement is critical in finding the root cause,  e.g., \textit{``You hit the shot and it was a winning rally, why? Because the opponent wasn’t there yet''} (C2).
Both C2 and C3 mentioned the inclusion of player poses to enhance the usefulness and engagement of the 3D game view. Therefore, we worked on player pose estimation after the third user testing.

During the user testing, we also elicited coaches' feedback on the VR environment.
% as all of them were first-time VR users. 
They agreed that VR provides additional benefits in analyzing a match video, such as immediate access to all relevant information, flexible viewpoints, and an interactive approach. C1 commented \textit{``it was very helpful to see the video and the bird going with the trajectory at the same time.''}
% \textit{``when I put the headset on I already have information that I may need without even having to watch the video''} \jui{this is vague. what did D1 mean?}. 
C2 moved to the bottom left of the court while watching the 3D game view because \textit{``this is where I sit as a coach''}. C3 shared that the interaction to select a shot and link to the actual match video is very helpful as \textit{``it’s important to know how the player put the pressure and create a situation [in the game]''}.

\subsection{Implementation}
% \re{We implemented VIRD interface using Unity3D~\cite{unity} and run on Meta Quest 2. CV algorithms~\cite{liu-2022, li-2022} were implemented in Python.
% Preprocessed match data were loaded and rendered onto the 3D scene at run time. VR interactions were implemented based on XR Interaction Toolkit\cite{xrinteraction}.
% VIRD interface is available at https://to-be-open.github.io.
% }
\re{
%The VIRB system combines a backend component and a front-end user interface. 
The VIRD backend is implemented in Pytorch and leverages CV models~\cite{liu-2022, li-2022} to extract data from badminton videos. 
The processed match data is subsequently rendered in real-time within the front-end's 3D scene.
The font-end of VIRD is built with Unity3D~\cite{unity},
including the user interface and the 3D scene.
To be compatible with the Meta Quest 2 platform,
we have implemented VR interactions using the XR Interaction Toolkit~\cite{xrinteraction},
ensuring a natural, intuitive user experience. 
The VIRD interface can be accessed at our public website: \url{https://github.com/ticahere/VIRD-demo}.
}
% \input{docs/@deprecated/5_VIRD@v1}
% !TEX root = ../main.tex

\section{Evaluation}
\label{sec:user-study}
% We designed three case studies to evaluate how well VIRD supports badminton experts analyze match videos. 
% We described the case studies and the computational performance of selected matches.

\subsection{Case Study Design}
\label{sec:case-study-design}
%Due to 
% the scarcity of the target users and 
%the complexity of analysis tasks, 
We conducted in-person case studies~\cite{lam_empirical_2012} with domain experts to evaluate VIRD on match analysis in four aspects: 1) data analysis method, 2) derived insights, 3) useful components, and 4) overall user experience.

\noindent
\textbf{Participants \& Data.} 
% Our study goal is to evaluate VIRD on match analysis in four aspects: 1) data analysis method, 2) derived insights, 3) useful components, and 4) overall user experiences. 
% 
We invited two high-performance coaches from the user testing phase (C1 \& C2; M=2; Age: 40-60) along with a US national team player (P1; M; Age: 20-25) who had been mentored by C1 for a decade. 
None of them had prior experience using VR outside of our study.
We selected three professional matches, including two public matches (M1, M2) and one personal match provided by P1 (M3).

\begin{itemize}
    \item M1: 2021 Denmark Open Final, MS, Momota vs. Axelson~\cite{match_axelson_momota}  
    \item M2: 2022 BWF World Champ. QF, WS, Yamaguchi vs. Marin~\cite{match_marin_yamaguchi} 
    \item M3: 2022 Mexican International, R32, MS, Ma (P1) vs. Castillo
\end{itemize}

\re{
We processed the match data as described in Sec.~\ref{sec:data_preprocessing}.
%We performed data preprocessing on these three matches using techniques described in Sec.~\ref{sec:data_preprocessing}.
}
% M1 was a quarter-final game between Marin and Yamaguchi in 2022 BWF Women's Single, lasting for 47 minutes. M2 was a final match between Axelson and Momota in 2021 Denmark Open Men's Single, lasting for 1 hour 33 minutes. M3 was a round 32 match between Ma and Castillo in 2022 XIII Mexican International Men's Single, lasting for 1 hour.

% [M1] 2021 Denmark Open - Final MS, Axelson vs. Momota
% [M2]  2022 BWF World Championship - QF WS, Marin vs. Yamaguchi
% [M3] 2022 Mexican International - R32 MS, Ma vs. Castillo

\para{Experiment Set-up.}
% The user study was conducted in a 300 sq ft meeting room. The participant wore Meta Quest 2 to use VIRD.
% We ran VIRD in Unity3D~\cite{unity} on a PC with a i7-11800H
% 2.30GHz processer and an NVIDIA GeForce RTX 3060 graphics card.
% We displayed VIRD on a Meta Quest 2 virtual reality headset with $1,920 \times
% 1,832$ resolution per eye and a 90 Hz refresh rate, connected to the PC through a 5m USB3 Type-C cable. 
% The VIRD view was also projected onto a 65" 4k TV screen connected to PC so the instructor can see the VR view. 
% 
The user study took place in a 300 sq ft meeting room, where participants used VIRD with a Meta Quest 2 headset. VIRD was run 
% from Unity3D~\cite{unity} 
on a PC equipped with an i7-11800H 2.30GHz processor and an NVIDIA GeForce RTX 3060 graphics card. The Meta Quest 2 VR headset has a resolution of 1,920 x 1,832 per eye and a 90 Hz refresh rate, connected to the PC via a 5m USB3 Type-C cable. The VIRD view was also projected onto a 65" 4K TV screen connected to the PC, allowing the instructor to observe the VR view.

\noindent
\textbf{Study Design.}
To evaluate how VIRD supports match analysis, 
each coach analyzed one public match for developing game strategy. C1 and C2 analyzed M1 and M2, respectively.
In addition, to evaluate how VIRD helps derive and communicate insights for coaching, 
both C1 and P1 analyze M3 in the same session.
% , we scheduled C1 and P1 together.   
During the study, C1 analyzed M3 using VIRD and provided coaching advice directly to P1, who watched C1's interaction on a TV screen.

\noindent
\textbf{Procedures.} 
We first introduced the study to the expert and obtained their consent to participate and be recorded. They agreed to disclose their identity in the paper. 
\re{
The experts first watched the match video on the desktop for 10 minutes to familiarize themselves with the players in the match, as they had not coached them before.
%The experts first warmed up by watching the match video on the desktop for 10 minutes. This step allowed coaches to familiarize themselves with the players in the match as they had not coached either player in the two chosen public matches before. % and did not apply to P1.
}
Next, we introduced key features of VIRD with a list of example tasks, such as \textit{``select all winners by Momota in G1''}, and asked the expert to explore the features freely. This training step took around 10 minutes. The expert then analyzed the assigned match for 10 minutes in think-aloud fashion. After match analysis with VIRD, they were asked to conclude their coaching advice. In addition, C1 performed another match analysis of M3 and shared his advice with P1 in the study for 10 minutes. Finally, we gathered feedback from the expert about their experience with VIRD in a post-study survey and a follow-up interview.  
Each study took 60 to 75 minutes and we compensated each participant with a \$50 gift card.

\noindent
\textbf{Measure \& Data Analysis.}
We recorded the user interaction, voices, and VR screen records for analysis. In the post-study survey, we collected subjective ratings on a five-point Likert Scale, including learnability, usability, usefulness of each feature, and overall satisfaction of VIRD. In the follow-up interview, experts commented on the most useful features, pros and cons, and suggestions for using VIRD in actual coaching.
To evaluate the experts' analysis and the insights they obtained, we performed text analysis on audio transcripts. We labeled user comments based on knowledge type, including prior knowledge, analysis, or insight. We also mapped user comments to VR screen records to extract the visualizations used in the analysis.

% \subsection{Computational Performance}
% We evaluated the performance of the shot and player pose detection of the three selected matches in the case study.  \jui{Do we need these? They just show that CLIFF performs well in our videos. Both MonoTrack and CLIFF are published work; I suggest we simply cite the papers and maybe provide a few numbers in 6.1 in 1-2 sentences to show how well they do on our games, and remove this section.}
% \begin{table}[h!]
% \small
% \begin{tabular}{|c|c|c|c|c|c|}
% \hline
% \textbf{Match} & \textbf{Length} & \textbf{\# Rally} & \textbf{Missing Frame} & \textbf{Total Frame} & \textbf{Accuracy} \\ \hline
% M1        &  93 mins & 109                        & 800                    & 56' 570              & \textbf{98,59\%}  \\ \hline
% M2        & 47 mins  & 73                         & 505                    & 24' 517              & \textbf{97,94\%}  \\ \hline
% M3         & 60 mins & 106                        & 2 976                  & 29 '086              & \textbf{89,77\%}  \\ \hline
% \textbf{Total} & 200 mins & \textbf{288}               & \textbf{4 281}         & \textbf{110 '173}    & \textbf{96,11\%}  \\ \hline
% \end{tabular}%
% \end{table}

% The missing frame column in the table corresponds to frames where at least one player was not detected by the algorithm. M1 and M2 have high accuracy ($>$97.9\%) with broadcast-quality video while M3,
% filmed with a phone camera, had approximately 90\% accuracy.
% The overall accuracy rate of over 96\% demonstrates the strong robustness and effectiveness of Cliff [] in detecting players. 

\subsection{Case Study Results}
We present the results of two case studies.
%, each with a different focus on the use of VIRD for match analysis. 
Case 1 examines a coach developing game strategies using VIRD. Case 2 explores a pair of coach and player
communicating insights for coaching. For both cases, we describe coaches' findings using the match player's last name.

\subsubsection{Case 1: Developing Game Strategy in a Match}

We demonstrate C2's analysis workflow on M2 with both desktop and VIRD, highlighting his analysis approach, insights, and interactions.

\begin{figure}[t!]
  \centering
  \includegraphics[width=\linewidth]{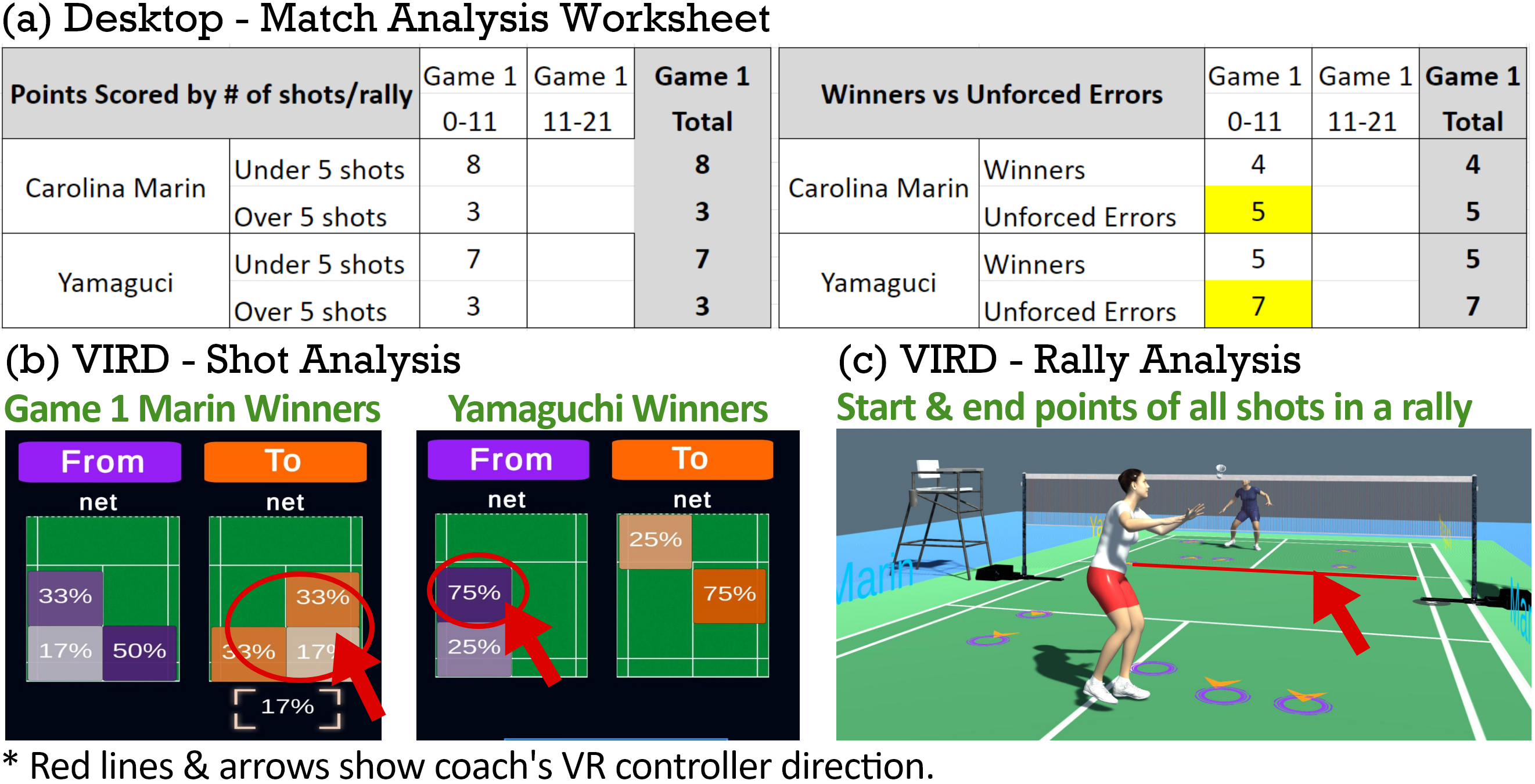}
  \vspace{-6.5mm}
  \caption{Visualizations used by the coach in Case 1. (a) Stats collected in a worksheet on the Desktop. (b) and (c) show the shot heatmap and 3D locations used to generate and present insights in VIRD.}
  \vspace{-6.5mm}
  \label{fig:case}
\end{figure}

 % 2022 BWF QF Yamaguchi vs. Marin
 \noindent
 \textbf{Desktop.} 
 During the 10-minute warm-up phase, the coach went through M2's first half (11 points) of game 1.
 \re{
  Using his usual video analysis approach, the coach went through the YouTube video sequentially, pausing or fast-forwarding to the rally, and manually recorded statistics (the number of winners, errors, and short rallies) on a spreadsheet after each rally (\autoref{fig:case}a).}
% Using his usual video analysis approach, the coach interacted with the video on YouTube by going through the video sequentially, pausing or fast-forwarding to the rally, and manually recording statistics 
 % \jui{can we use statistics throughout the paper? stats is not formal use}
% (the number of winners, errors, and short rallies) on a spreadsheet after each rally, as shown in \autoref{fig:case}a. 
 % analsyis
% Upon analysis, he focused on 
Using the collected stats and observations in the video, he discovered that Marin had won 8 out of 11 points very quickly, with 4 winners versus 5 unforced errors.
 Further, he observed that Marin was playing very flat and trying to push the tempo, leading to Yamaguchi only playing from a small area on the court. 
 
 % insights
 These observations led to two coaching insights.  The coach stressed that these were initial observations that he would usually first validate in more detail. 
 % With the limited time, the coach stressed that he points out the trend but would cross-reference for validations.\jui{this sentence seems a bit out of place. logically it will connect better if you say right away what are the two insights.}
 First, Marin is playing very fast and not moving the opponent. 
 %With Yamaguchi barely moving out of the box, 
 \textit{``[Yamaguchi]'s getting more comfortable with what Marin is doing.''} The advice is that Marin needs to utilize the backcourt and open the court more. To explain this insight, the coach pointed to the mid-court areas on a court diagram.
 Second, Marin is pressing the match and only playing flat shots. \textit{``These are world-class players. You can't just do the same thing the whole time.''} The coach thinks she needs to change techniques, such as varying the speed and angle of the shot. % shorten, or use more fake shots.
Lastly, the coach commented that 5 unforced errors were a little too high for the first half of the game. Yamaguchi is not moving much, leading Marin to waste her energy, \textit{``Marin's going to make more mistakes in the long-term if she doesn't change the strategy.''}

\noindent
\textbf{VIRD.} 
During the 10-minute match analysis with VIRD, the coach used the filters to focus on each player's winners and errors separately. He also drilled down to specific rallies or the shot video to verify observations, and focused on spatial aspects such as shot location, distribution, and shot trajectory in the analysis.
% analysis
To continue his analysis of M2 from the warm-up, he selected the first half of game 1 and explored the shot distribution by players and shot outcomes.
% \jui{he actually studied the same portion of the game as in desktop?}. 
He found that Marin's winners came mostly from the back and she attacked the bottom right corner (\autoref{fig:case}b left), while most of Yamaguchi's winning shots were from the front (\autoref{fig:case}b right). 
%Looking at errors, 
He also found that Marin's errors were pretty evenly distributed while Yamaguchi had more errors in the front. 
After an overview, the coach continued his analysis based on different hypotheses, such as wanting to see how Marin did on her winners because that's an important part of her game. He went through each of the winner rallies in detail and examined the shot locations on the court. He commented \textit{``Look at all the dots on the Yamaguchi's court, none of them pass this white line back here''}, pointing at the court with the VR controller (\autoref{fig:case}c).
He further focused on short rallies (less than 10 shots) using the Rally Menu, and observed that Marin's backhand serve was really flat, giving Yamaguchi scoring opportunities.

% insights
The coach was able to verify his previous insights obtained on the desktop with concrete evidence while pointing out additional details.
% 1st insight: not moving the opponenet
First, he referred to the shot locations on the court (\autoref{fig:case}c) to demonstrate that Marin is barely moving the opponent. Further, based on the heatmap showing Yamaguchi has most of her winning shots from the front (\autoref{fig:case}b right), he suggested
\textit{``Marin should try to avoid this corner because Yamaguchi is creating scoring opportunities from this corner.''} 
% 2nd insight: flat shot
Second, he replayed a short rally and pointed out the flat shot, \textit{``Marin's backhand serve was really flat. [Yamaguchi] didn't have to move at all. She scored right away''}. Based on examining the short rallies, the coach suggested that Marin should
\textit{``either make the serve higher ... or don't use that type of serve because it's not working.''}
 \re{
 To explain these findings, the coach used actual video clips and spatial data and compared patterns between players and rallies. 
 %To explain these findings, the coach used actual video clips and spatial data to showcase the results, and compared patterns between players and rallies to verify observations. 
 }
 Throughout the analysis, the coach also used the first-person view to describe findings, such as \textit{``I want to avoid this corner and play the other ones.''}

% \jui{Should we add a short summary paragraph here to strengthen the argument of using VIRD? It seems that the message is "what C2 can do on desktop, he can do in VIRD. In addition, using VIRD he can do X, Y, Z better/more efficiently.} 

\noindent
\textbf{Case 1 Summary.} 
\re{
C2 analyzed the first half of game 1 in match M2 with Desktop and VIRD. The coach effectively verified two initial observations from Desktop using  VIRD and explained his insights with spatial data visualizations and specific rallies. He also seamlessly iterated between summary data and detailed game views to support his analytic reasoning.
}
%In Case 1, C2 analyzed the same match duration of M2 (first half of game 1) with Desktop and VIRD. The coach effectively verified two initial observations from Desktop using  VIRD, including that Marin was not moving the opponent and her shots were too flat, and explained his insights with spatial data visualizations and specific rallies. He also seamlessly navigated between summary data and detailed game views to support his analytic reasoning and iteration.

\subsubsection{Case 2: Verifying and Communicating Coaching Insights}
We describe how C1 verified and shared coaching advice with P1 using VIRD and how P1, as a player, obtained analysis insights. % from the player's perspective. %from the coach's perspective

% \jui{This section has two main paragraphs: Coach and Player. Before going into these paragraphs, you should explain what they are and why they are structured like these}

\noindent
\textbf{Coach.} Prior to our study, C1 had already spent 2 hours analyzing M3 and shared his insights with P1 virtually. 
\re{He had asked P1 questions about his strategy and his opinion about the match.}
%He coached P1 by asking questions about his strategy and his own understanding and opinions about the match. 
Based on the discussion, the coach followed up with match insights and statistical trends to explain his advice. % shorten They did not get into specific rallies in the actual video due to limited time.
  In this case study, the coach used VIRD to explore the match data in more depth, to communicate directly with P1, and to verify previous insights and update some original hypotheses.
  
The match between Castillo and Ma was won by Ma  (P1) with 21-11 (G1), 19-21 (G2), and 13-21 (G3). 
% \jui{repeated and colliding use of G1, G2, G3} 
The coach first read the score of the match from the Match Summary, and selected G2 with a score of 19-21 won by Ma because it was a close game. He focused on 18 errors made by Ma and examined each rally. Utilizing the virtual red shot arc to immediately pinpoint where the errors were coming from in each rally, he quickly browsed three rallies of errors and identified that Ma made all mistakes on defensive shots. 
This confirmed one of his previous analysis insights, where he pointed out P1's recovery shots gave the opponent too many opportunities, and suggested P1 working on his defense, 
\textit{``this shows we're working on the right thing because he's making mistakes here''}.
To demonstrate this insight, he continued to select another rally and used the VR pointer to point at the red shot arc, \textit{``you can see the mistake from this mid court when his opponent attacks.''} Based on this analysis, he asked follow-up questions to drill down to the root cause with the player, such as footwork issues. %or physical fitness issues.  

The coach analyzed the 14 winners by Ma and found that 43\% of the winners were hit to the front left area on the heatmap. He filtered the shot location to focus on the 6 winners hit to the front left. He was surprised, since P1 is not confident in their net play. He pointed to the heatmap to show P1 that \textit{``your front is better than what you expected. You scored from the front the most''}.  He ended the analysis with a comment that he would go through each rally with the player in more depth, e.g., \textit{``Did you score because you’re always in the front, or was it your skill?''}

\para{Player.} 
The case study with P1 focused on how VIRD helps players analyze their own match in an immersive environment.
%shorten and improves their understanding on top of their existing knowledge.

After being introduced to the features of VIRD, the player analyzed M3 for 10 minutes. He first looked at winners and focused on comparing his shot locations and trajectories across each game. He found that his winners came majority from the back left, \textit{``the first game was 60\% from the back left''}. Further, he found almost all of his winners were going down or flat across three games by examining the shot arcs on the virtual court. 
He then confirmed his observations with the rally video.
%drilled down to the rally video to confirm the observations. 
By comparing data across games, he found that he had more winners in the front but not in the back in G2, but upon further investigation of a rally in G2, \textit{``for that rally it seemed like I did win it in the front but a lot of it was set up in the back actually''}.
\re{
Looking at the winner shot analysis, he was surprised that the majority of winners came from his backhand side, \textit{``I thought for me it was a lot easier in general to attack from the forehand side''}.
%he found that the majority of winners coming from his backhand side was surprising to him as \textit{``I thought for me it was a lot easier in general to attack from the forehand side''}. 
He also contemplated that he might need to find more ways to keep the shot down as it seemed beneficial if he did not hit it up as often.}

The player continued to analyze his errors in each game and compared the shot heat map. Seeing too many errors shown in a game, he chose to split each game into half. 
He found that most of his errors came from the back across all games. He moved in the virtual court to match his position on the court from the first-person view and pointed at the shot arcs and heat map on the virtual court to communicate his observations. He found his errors were mostly on the backhand side in the second half of G2
% \jui{undefined terminology?}
, while the majority were on the forehand side. %shorten (five out of six half game sets). 
He contemplated that \textit{``maybe he started changing strategy...''} as G2 had a tight score (19-21) where the opponent was close to winning the match.
Upon analyzing the shots on the virtual court and previewing them in the video view, he also found that all of the errors were very high arcing, \textit{``the arc tells me I'm losing because of defensive shots. It's either because I'm hitting into the back or my shots in the front are too high''}.
%, reflecting the coach's advice.
%To conclude, 
The player commented he would use this tool to go through every rally, and considered VIRD very helpful to get a sense of his overall performance and identify areas to discuss with the coach. 

\noindent
\textbf{Case 2 Summary.} In Case 2, C1 analyzed M3 and shared coaching insights with P1, while P1 also analyzed his own performance. The coach used VIRD to verify one previous insight (P1 needs to improve defensive shots) and discovered a new insight (P1 has better net shots than he thought), and used a combination of spatial visualizations and rally videos to explain his insights to the player. The player focused on finding patterns in his performance with spatial visualizations and drew conclusions with two new findings about his shots, including most winners coming from his backhand side and downwards, and most errors coming from the back across all games.

\begin{figure}[t!]
  \centering
  \includegraphics[width=\linewidth]{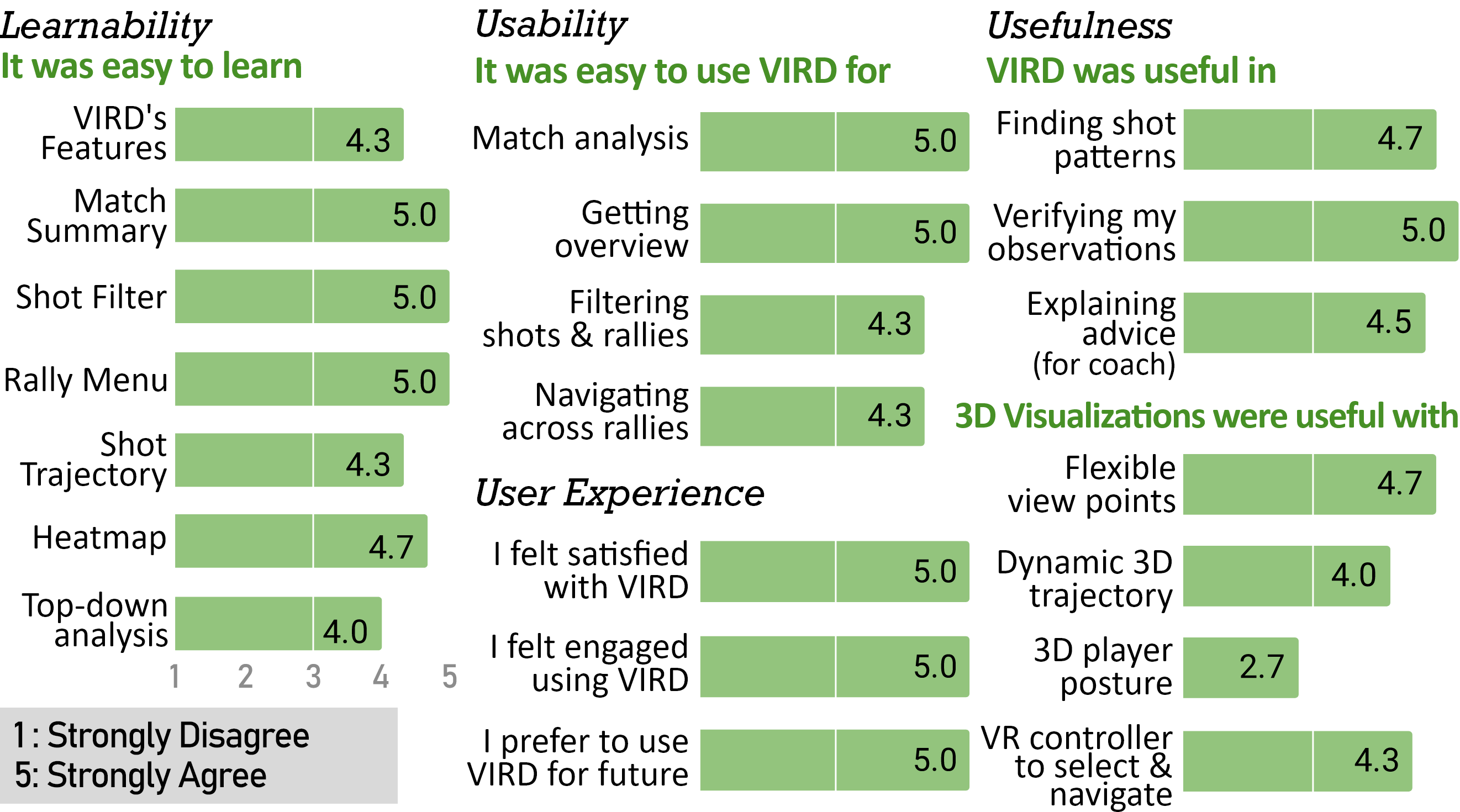}
  \vspace{-5mm}
  \caption{Subjective ratings from experts (N=3) analyzing matches with VIRD show high learnability, usability, and usefulness of VIRD. 3D player posture is rated less useful due to technical limitations. Experts were satisfied, engaged, and preferred to use VIRD for match analysis.}
  \vspace{-4mm}
  \label{fig:rating}
\end{figure}

\subsubsection{Post-study Survey \& Interview}
\autoref{fig:rating} shows the average subjective ratings collected in the post-study survey, ranging from 1 (strongly disagree) to 5 (strongly agree).
%The average subjective ratings of the post-study survey are shown in \autoref{fig:rating}, ranging from 1 (strongly disagree) to 5 (strongly agree).
Overall, experts rated VIRD with high learnability, usability and usefulness ($\mu \geq$ 4.0 except 3D player posture). % shorten We discussed each item with qualitative feedback from the follow-up interview.

\para{Learnability.} Experts found it easy to learn the features in VIRD. Particularly, the data provided in each visual element (Match Summary, Shot Filter, Rally Menu) were clearly understandable. The overall top-down analysis method, and spatial data visualizations in shot trajectories and heat maps were also properly learned during training. 
%shorten: Given the short study, coaches also expressed they will be able to use the tool to more extent with practices.

\para{Usability.} Experts rated the ease of use of VIRD for analyzing matches highly in each of the analysis tasks, including getting overviews, filtering shots and rallies, and navigating across rallies.
The main issues arose in operating the VR controllers, such as clicking on the trigger button. Some suggestions included adding an onboarding tutorial to train users on accurately interacting with each component.   

\para{Usefulness.} 
Experts found VIRD helpful for match analysis from finding shot patterns, verifying insights, and explaining coaching advice. The coaches found the static data panels (Match Summary \& Shot Filter) most helpful as they provide the foundation of analysis to link to videos and spatial data visualizations, and visually showcase data and videos to the player. 
Further, experts found it helpful to have a 3D virtual court with flexible view points, use a VR controller to select and navigate, and view dynamic 3D shot trajectories. 
The interactive approach was highlighted by both the coach and player as a benefit of VIRD, which supports an iterative analysis loop as well as linking static data to dynamic video and shots. 
The player found 3D visualizations (heatmap and trajectory) very useful in finding shot patterns and generating insights on his performance, especially being able to compare the video with 3D game from different angles.
However, 3D player posture was considered less helpful ($\mu$=2.7) as 
coaches found the players off balance occasionally due to technical limitations. Experts mentioned they found the player positions in 3D game views helpful, and did not pay much attention to the actual posture. We observed that coaches were mostly interested in player movement and whether they were in good positions when hitting the shot. According to experts, observing slightly off-balanced player postures did not significantly impede their ability to comprehend the 3D gameplay. In cases where it was necessary, they would refer to video views for comparison.
% \jui{Should we expand on this a bit more given it is the lowest score by far? Why is player's pose off-balance a bad thing? What about their positioning? One motivation for having pose is their on-court position. Did that help? The other is for the pose when players hit the shots (e.g., in backhand backcourt). Did it help for those cases?} 
% In some occasions, coaches found the player poses off balance and had to look at the video to confirm. 
% Although pose and shot detection models are not the main contribution of our research, the accuracy is crucial to support a complete match analysis workflow.
% We discuss the limitations and opportunities of computer vision for sports further in Sec.~\ref{sec:top-down-analysis}.

\para{User Experience.} Experts felt satisfied, engaged, and prefer to use VIRD for analyzing match videos (all $\mu$=5.0). 
% pros
The major advantages for coaches were getting instant access to data, using an interactive approach, and access to 3D visualizations, which could lead to a huge reduction on the time to perform match analysis (less than 30 minutes vs. 3-5 hours). For the player, the pros are having spatial data to help dive into one's own strengths and weaknesses.
% cons
On the downside, coaches found the shot and pose detection occasionally inaccurate. Although they could refer to the actual video to verify, it hindered the experience of viewing the entire game in 3D. The player felt that the VR environment made him feel like he was in a game and he would get distracted. %shorten,  \textit{``there's so many things I want to do or play stuff''}.
As a player, he considered using this system for game watching instead of analyzing, since \textit{``If I'm training a lot and really tired, then going into this [VR], of course I want to have fun''}.
% If I'm training a lot and also competing, I'm really tired. Then going into this [VR], of course I want to have fun
% !TEX root = ../main.tex

\section{Discussion}
\subsection{Top-down Analysis Approach for Sports Videos}
% \subsection{Data-Driven Video Analysis without Losing Contexts}
\label{sec:top-down-analysis}

% 1. top down approach is preferred by experts as it shows summary and provides data to support detailed analysis
% 2. however, it is not prevalence for two reasons. (1) to collect high-quality data, abundance resources is required. Only a fragment of athletes have access to it, like NBA league, national teams of popular sports. (2) with fewer resources, no data experts can provide good analysis and good data. Therefore, sports experts do not trust the data; or have to manually collect the data, like coaches in our study.
%  3. Our design has shown that, by maintaining the contextual understanding of the data, sports experts can use top-down approach to analyze game, without sacraficing their domain knowledge nor compromising their coaching time.
% therefore, it is important to provide context during the analysis.

% \jui{This section is not very focused. It seems at first about the top-down approach we designed and how it is preferred, but the last paragraph is for combining CV and human-in-the-loop, and the mix use of video and reconstruction. To make it worse, the section title is ambiguous. Make up your mind on what you want to say here and just say it, with the proper title.}

We proposed a top-down data analysis approach in our study to enhance badminton coaches' video analysis workflow. 
The approach involved providing an overview of the match and then using filters and visualizations to narrow down the area of interest and identify specific game moments.
% The approach involved coaches first getting an overview of the match and then using filters and visualizations to narrow down the area of interest and drill down to specific game moments. 
 All experts agreed that this approach was effective in supporting coaches to generate and verify insights in the case study.
% All experts in our study favored this approach as it supported coaches in generating and verifying insights effectively in the case study. 

While the top-down analysis workflow is a well-established visual analytics approach~\cite{shneiderman2003eyes}, 
% described in \textit{``Overview first, zooming and filtering, details on demand''}~\cite{shneiderman2003eyes}, 
it is rarely used by sports domain experts for analyzing sports videos. 
Instead, sports professionals, such as scouts and coaches, typically watch individual game videos to evaluate a player's performance 
due to the inaccessibility of high-quality data from videos, and limited resources that lead to experts relying on their own interpretation and annotation of videos.
% There are two main reasons for this. First, collecting high-quality data from videos requires advanced sensing technologies and input from data experts, making it inaccessible to most sports, e.g., no player tracking data is available for professional badminton matches. 
% Second, sports experts often rely on their own annotation and interpretation of videos due to limited resources.  This can lead to a gap between data and context during the analysis, resulting in less comprehensive and communicable insights, as found in Sec.~\ref{sec:gaps}.  
However, this approach can result in a gap between data and context, leading to less comprehensive and communicable insights, as found in Sec.~\ref{sec:gap3}. 

To tackle this problem, we combined two promising avenues for sports analytics: CV-based data collection and human-in-the-loop analysis.
% with a direct contextual understanding of the data. 
Our study found that experts leveraged immediate access to both summary data and video to perform top-down analysis,
and integrate multiple data sources to develop and communicate their insights, such as domain knowledge, static and dynamic data. 
% to develop coaching advice and communicate with the players. 
Even when automation fails, experts can use the actual video to verify the data. As computer vision techniques continue to improve, we envision automatic data collection to benefit more sports domains,
% that more sports can benefit from data automation, 
while human-centered design empowers experts with an effective top-down analysis approach without losing contextual understanding of the data.
% empowering experts to perform top-down analysis for sports videos with a direct contextual understanding of the data.

% While computer vision can automatically detect sports events and statistics at an increasingly high accuracy~\cite{}, 
% in a highly personalized and collaborative task like sports coaching, expert knowledge will still be the dominating factor in the analysis process.
% Therefore, it is important to maintain the underlying context to allow experts performing analysis in depth while saving time on the low-level tasks. 
% In our study, we found that experts leveraged the immediate access to both static data and video to perform top-down analysis, integrating multiple data sources in their insights (expert knowledge, static data, dynamic data) and developing a coaching plan to communicate with player. 
% Even when the automation fails (e.g., inaccurate player pose), experts were able to use the actual video to verify. As CV and ML techniques continue to improve, we envision more sports can benefit from the data automation and empower experts to perform data-driven video analysis without losing contexts. 

% \textbf{Human-in-the-loop analysis vs. Data-in-the-hand analysis}

\subsection{Immersive Video Analysis for Sports Coaching}
% benefits of VR for sports analytics

Based on our case studies, we found that an end-to-end immersive analytic pipeline like VIRD can be suitable for sports coaching in badminton.
% \jui{personally, I am not too fond of excessive "bolding" everywhere. The section title is bold; the paragraph start is bold, and the bullet points are bold. This context switching makes it harder to get the structure of the article from a glance.}
Both coaches and the player in the studies were able to achieve their match analysis goals throughout the entire analysis pipeline in VR, from data exploration, insight generation to communication (using VIRD to showcase their insights to viewers on a TV screen in our study). 
% \jui{exactly how does communication happen in VR? Maybe I missed it...}. 
Unlike immersive analytics for scientific data analysis, where several analytic steps such as analyzing abstract data are better conducted in traditional desktop environments~\cite{bach2019immersive, hubenschmid2022relive}, sports data are intrinsically spatial and dynamic, making analyzing and presenting insights using videos and visualizations in 3D desirable. Furthermore, analyzing sports videos for coaching relies heavily on domain knowledge without the need for complex data manipulation. Thus, we found immersive analytics provide several advantages for sports coaching.

\para{1. Situated visualization reduces context-switching costs and shortens the path from hypotheses to insights.}
As both coaches commented, the most beneficial feature of VIRD was the immediate access to all the required data, \textit{``when I put the headset on I already have information that I may need without even having to watch the video first''} (C1).
In the standard workflow on the desktop, coaches spend much time navigating and finding the critical game moments (e.g., when an error happens) while tallying rallies of interest on a separate note. This process induces high context-switching costs~\cite{wang2000guidelines} as the coaches need to constantly re-interpret the changed views between data in the notes and the game moments in the video, leading to a longer analysis cycle.  
% \jui{this seems to be a hypothesis. how did we confirm?}.
With large screen space and situated visualization placed in context (e.g., shots on the badminton court), coaches are presented with all required data in multiple views with spatial-continuous movement, which was found to reduce context-switching costs~\cite{yang_embodied_2021, plumlee_zooming_2002}. 
Therefore, experts can leverage their visual working memory~\cite{plumlee_zooming_2002} and focus exclusively on the match analysis.
% which Plumlee et al. described in their predictive model that multiple views are more efficient for tasks requiring high visual memory compared to zooming interface. 
% \jui{what does "visual working memory" mean? Clarify, cite, explain when appropriate.}.
This was shown in the hypothesis-driven workflow with VIRD, where experts can plan their analysis and immediately verify their insights with data, \textit{``I'd like to see where I made a mistake on the court. Almost all of it was on the forehand''} (P1).

% \textbf{2. Embodied interaction improves visibility of critical game moments.}
\para{2. Multi-modal data analysis improves visibility of critical game moments.}
An essential task in video coaching is identifying critical game moments to reveal root causes, \textit{``when we coach a player, we have to pinpoint the exact cause and outcome''} (C2).  
% These critical moments were identified by experts based on scores (e.g, last few rallies in a tied game), duration (e.g., short rally), and when winners and errors happened.
The ability to breakdown a rally stroke by stroke and instantly preview each shot in the video with embodied interaction in VR allows experts to directly access and focus on these critical moments.
One coach even suggested \textit{``it would be good if there's a way to play a loop of all the winning shots. Just because I'll spend a lot of time between going into different strokes.''} 
Beyond individual shots, experts also dive into selected rallies in further detail. By comparing static data and 3D visualization from the game, they can develop a more comprehensive analysis from data summary to key moments. For instance, the coach (C1) went through all error rallies and pointed out the player's weakness on defensive shots as he pinpointed the location of error shots on the court with the VR pointer.

% Experts examined player movement, reactions, shot sequence in the rally to determine the cause and develop strategy to improve the performance. 
\para{3. Immersive 3D visualizations deepen game understanding and engagement.} 
Experts expressed excitement that they could move freely in the virtual court and watch the game from different angles. Further, all of them were excited seeing the moving bird and the 3D reconstructed game.
% commenting with surprise like \textit{``That’s pretty cool!''} 
Throughout the analysis, experts felt engaged and interactive, \textit{``I like how I can move my body around and face the shot. I find that more beneficial than just looking at a TV screen or computer''} (C1). 
Experts also expressed their tendency to view a match in 3D and refer to the video only when the 3D view did not make sense. A coach also suggested adding rackets to improve the 3D view.
% The player felt it was fun to just rewatching himself in 3D and compare it with the video.
Using VIRD, experts obtained deeper insights
% \jui{can you provide an example?} 
on the spatial aspects in their analysis. P1 observed most of his winners were hit downwards from his backhand side from analyzing the shot locations and trajectories. 
  % as 3D visualizations provides shot breakdown and objective viewpoints
%However, with the errors in shot and pose detection, experts expressed they had to refer to the video to compare actions in the actual game. A coach suggested fixing the postures and adding rackets to improve the 3D view. 
% 
One interesting finding was the sense of presence in VIRD.
A coach used the first-person view to describe his analysis for a player in the match, saying \textit{``these are my errors''}, while the other coach moved to the coaching position by the court and the player moved to his side of the court to view the game from a first-person viewpoint.

\subsection{Limitations \& Generalizability}
\textbf{Limitations.} 
% shorten: We discuss the limitations of automatic data collection accuracy and the small number of domain experts in our study. 
% Further, we envision extending our work in the direction of remote collaboration, natural language-based interaction, and leisure game viewing.
% \jui{I will start the section by a paragraph summarizing/enumerating the limitations, before jumping into each of them in order of importance.}
% 
Our computer vision models are around 90\% and 96\% accurate in detecting shots and player poses, which causes confusion when inaccuracy occurs. While experts can still perform match analysis by accessing the video view in VIRD, this was mentioned as an area for improvement by all experts. As CV techniques advance, we envision the limitation on automatic data collection can be largely improved with better-trained models, making our approach more reliable. 

\re{Due to the limited access to high-performance badminton experts, such as Olympian coaches and players,}
\re{our study reports the feedback from a few domain experts. 
% we were fortunate to get connected with. 
We believe the identified problem is significant and common among badminton athletes, but our solution might not generalize to all experts due to varying analysis approaches and resources among coaches and players.} Instead, we consider our main contribution to be a design study exploring the use of immersive analytics in real-world sports coaching.

% generalizability
\noindent
\textbf{Generalizability.} We believe our established data preprocessing pipeline (based on MonoTrack~\cite{liu-2022} and CLIFF~\cite{li-2022}) and the immersive and interactive way to analyze multi-modal game data in badminton can be applied to general match videos and
benefit the broader community beyond professional coaches, such as players at all levels, and other racket sports.
With the VR benefits in visualizing spatial data and revisiting critical game moments, we also envision expanding VIRD beyond match analysis for leisure game viewing or broadcasting, as noted by the player that watching the game in 3D was fun.

% \noindent
% \textbf{Future Work.} Some exciting future work includes supporting remote collaboration between the coach and player in a shared immersive space, which can address the gap of limited coaching time and lack of support for video discussions found in the formative study. 
% We also foresee integrating natural language as an input method to further lower the context-switching costs throughout match analysis, as using VR controllers to select and filter data was found clunky for first-time users. 

% shorten: For example, coaches wanted to filter all winners shot won by Marin in game 1; Instead of clicking three buttons on the UI, this instruction can be parsed automatically by a large language model such as GPT~\cite{brown2020language}.
% 

% \jui{I would stress that many parts of offline preprocessing and VIRD can be repurpose for exactly this.}.

% \subsection{Implications for Sports Analytics in XR}
% % technology & research
% - computer vision for game reconstruction and analysis
% >> it is getting much better with video training data

% - XR technologies 
% >> technology is getting better and more available

% - human-AI interaction, personalization
% >> expert in the loop

% - future work: remote collaboration, game viewing for athletes and fans, simulated training (AR)

% !TEX root = ../main.tex

\section{Conclusions and Future Work}
In this study, we introduced VIRD, an immersive badminton match video analysis tool for high-performance coaching based on a formative study with Olympic coaches and players. 
VIRD employs a top-down analytic approach in VR with 3D reconstructed game views and multi-modal data analysis.
Experts successfully developed game strategies and effectively communicated insights using VIRD in case studies, showcasing the advantages of immersive analytics in badminton coaching. 
These benefits include reduced context-switching costs, enhanced visibility of critical game moments, and a deeper understanding of and engagement with the game through situated 3D visualizations.

Promising future work includes enabling remote collaboration between coaches and players in a shared immersive VR space, addressing the limited coaching time and insufficient support for video discussions. Additionally, VR could be employed to simulate game scenarios for enhanced athlete training.  
Incorporating natural language input methods, such as GPT, may also help minimize context-switching costs during match analysis, enabling more efficient analytical iteration.

% Results suggest that when presented with video context, CV-based data collection can support
% effective top-down analysis for sports coaching.
% % human-in-the-loop analysis for sports game analysis. 
% Experts expressed high satisfaction 
% analyzing match videos with VIRD. 
% % compared to current methods. 
% Further, immersive analytics show benefits for sports coaching in presenting data in-situ to reduce context-switching costs, improving the visibility of critical game moments with embodied interaction, and deepening game understanding and engagement with immersive 3D visualizations.

% !TEX root = ../main.tex

\section*{Supplemental Materials}

Our supplemental materials, including post-study survey questions for the case study, are available on OSF at \url{https://osf.io/92px8}. The front-end code for VIRD’s VR interface described in Sec.\ref{sec:vird} has been open-sourced and can be found at \url{https://github.com/ticahere/VIRD-demo}.
Our other code and video data are the intellectual property of Adobe Research and Badminton World Federation.

\acknowledgments{
This work is supported by Adobe Research, NSF grants III-2107328 and IIS-1901030.
We are grateful for the invaluable support of Andy Chong, Daphne Chang, Kai-Hsin Chang, Justin Ma, Hsiao-Ma Pai, Pashupati Paneru, Raju Rai, Rena Wang, Paul Liu, and Toby Ng. Their dedication to advancing badminton has significantly inspired and contributed to this project.
We thank the reviewers for their valuable comments.
}

% \clearpage
% The authors wish to thank A, B, and C. This work was supported in part by
% a grant from XYZ (\# 12345-67890).}

% \bibliographystyle{abbrv}
\bibliographystyle{abbrv-doi-hyperref}
%%use following if all content of bibtex file should be shown
%\nocite{*}
\bibliography{references}
\end{document}